\documentclass[twocolumn,preprintnumbers,amsmath,amssymb,superscriptaddress]{revtex4-2}
\UseRawInputEncoding

\usepackage{latexsym,amssymb,amsthm,amsmath,epsfig,braket}
\usepackage[left=2.00cm, right=2.00cm, top=2.00cm, bottom=2.00cm]{geometry}
\usepackage{xcolor}
\usepackage{svg}
\usepackage[colorlinks, citecolor={blue!80!black}, urlcolor={blue!80!black}, linkcolor={red!50!black}]{hyperref}
\usepackage{float}
\usepackage{hyperref}
\hypersetup{
    citecolor=red,
    colorlinks=true,
    linkcolor=blue,
    filecolor=blue,      
    urlcolor=blue,
}

\begin{document}
\title{Quantum magnetotransport in monolayer $\mathrm{Pt_{2}HgSe_{3}}$}
\author{Muzamil Shah}
\affiliation{Department of Physics, Quaid-I-Azam University Islamabad, 45320, Pakistan}
\author{Imtiaz Khan}
\email{ikhan@phys.qau.edu.pk}
\affiliation{Department of Physics, Zhejiang Normal University, Jinhua, Zhejiang 321004, China}
\affiliation{Zhejiang Institute of Photoelectronics, Jinhua, Zhejiang 321004, China}
\author{Kashif Sabeeh}
\email{ksabeeh@qau.edu.pk}
\affiliation{Department of Physics, Quaid-I-Azam University Islamabad, 45320, Pakistan}
\author{Muhammad Sabieh Anwar}
\email{sabieh@lums.edu.pk}
\affiliation{Department of Physics, Syed Babar Ali School of Science and Engineering, Lahore University of Management Sciences (LUMS), Opposite Sector U, D.H.A., Lahore 54792, Pakistan }
\author{Reza Asgari}
\email{asgari@theory.ipm.ac.ir}
\affiliation{School of Quantum Physics and Matter, Institute for Research in Fundamental Sciences (IPM), Tehran 19395-5531, Iran}
\date{\today}

\begin{abstract}
We present a theoretical framework to investigate quantum magnetotransport in monolayer jacutingaite, focusing on its response to external electric fields and off-resonant circularly polarized laser irradiation. Our analysis reveals a sequence of topological phase transitions triggered by tuning these external parameters. We find that the zeroth LL exhibits spin- and valley-polarized splitting, leading to four distinct peaks in the DOSs for the $K$ and $K'$ valleys. Using the Kubo formalism, we calculate both longitudinal and Hall magneto-optical conductivities based on the Kane-Mele model. Our results demonstrate that external electric, magnetic, and off-resonant optical fields can control these conductivities.
These findings highlight monolayer jacutingaite as a highly tunable platform with strong potential for future applications in photonics, optoelectronics, and topological quantum devices.

\end{abstract}

\maketitle

\

\section{Introduction}
Due to its remarkable electronic, mechanical, optical, and magneto-optical capabilities, graphene's discovery \cite{novoselov2004electric} sparked a wave of interest in two-dimensional (2D) materials. Because of these qualities, graphene is becoming a leading material for optoelectronic applications of the future. Since graphene's breakthrough, various 2D hexagonal materials have gained attention due to their unique topological properties. These include buckled Xene monolayers, which include silicene, germanene, and stanene \cite{ezawa2012topological,ezawa2013spin,Bampoulis202023quantum}. These systems have created new opportunities in spintronics and valleytronics because of the link between spin and valley degrees of freedom \cite{PhysRevB.87.155415,wang2015silicene}.

However, the relatively weak spin-orbit coupling (SOC) and centrosymmetric nature of buckled Xenes limit their effectiveness in certain quantum phenomena. In contrast, materials such as gapped graphene and transition metal dichalcogenides (TMDCs), which possess broken inversion symmetry, provide more versatile platforms for manipulating spin- and valley-dependent properties, making them promising candidates for advanced spintronic and valleytronic devices \cite{pesin2012spin,schaibley2016valleytronics}.

The first large-gap quantum spin Hall insulators controlled by the Kane-Mele model have been found theoretically as monolayer jacutingaite family materials ($\mathrm{M_{2}NX_{3}}$) \cite{marrazzo2018prediction}. The presence of the Kane-Mele phase in monolayer $\mathrm{Pt_{2}HgSe_{3}}$ and its counterpart in bulk crystals has been confirmed by experimental observations using scanning tunneling microscopy and angle-resolved photoemission spectroscopy  \cite{kandrai2020signature,cucchi2020bulk}. Because of their exceptional structural stability and topological characteristics, jacutingaite materials are very promising for use in quantum technologies and optoelectronic devices in the future.

When a two-dimensional electron gas (2DEG) is subjected to a perpendicular magnetic field, its continuous energy spectrum becomes discretized into a series of quantized LLs (LLs)—a hallmark of the quantum Hall effect \cite{klitzing1986quantized}. These discrete LLs are central to understanding various magneto-transport and magneto-optical phenomena, such as the quantum Hall effect, cyclotron resonance, and magneto-optical absorption \cite{stern1967properties,klitzing1986quantized,novoselov2005two,zhang2005experimental,sadowski2006landau}.

In recent years, the formation and behavior of LLs in two-dimensional materials and topological insulators under magnetic fields have been widely explored \cite{chu2014valley,yuan2018chiral,li2013magneto}. In graphene, the energy dispersion of LLs follows a $\sqrt{nB}$ dependence where $n$ is the LL index and BB is the magnetic field strength—leading to unequally spaced levels \cite{gusynin2007anomalous}. The conduction and valence bands in graphene are symmetric, with a flat zeroth LL located at zero energy. Optical transitions between LLs give rise to magneto-optical conductivity spectra, which are effective instruments for examining electronic characteristics like Fermi velocity, band gaps, and band structure features \cite{ando1982electronic,giuliani2005quantum,castro2009electronic,goerbig2011electronic,tabert2013valley}. Rich information on the underlying quantum states may be found in the resonance peaks in these spectra, which correspond to permitted inter-LL transitions. Graphene, silicene, $\mathrm{MoS_2}$, and phosphorene are among the 2D materials whose magneto-optical (MO) responses have been thoroughly studied over the last ten years \cite{castro2009electronic,tabert2013magneto,shah2019magneto,chu2014valley,tahir2015magneto, PhysRevB.107.235417}. These investigations have shown promise for advanced optoelectronic and quantum technologies.

In 2D quantum materials, the band gap may be efficiently tuned by a variety of external stimuli, including electric fields \cite{ezawa2012spin,yokoyama2013controllable} and antiferromagnetic exchange fields \cite{hajati2016valley}. Specifically, a special platform for investigating non-equilibrium events in these systems is provided by light-induced quantum effects \cite{bao2022light,kibis2010metal}. Both theoretical and experimental investigations have shown that a transition from a metallic to a band-insulating phase may be induced by the interaction of off-resonant circularly polarised light with massless Dirac fermions in graphene \cite{bao2022light,kibis2010metal}. Such effects in 2D Dirac materials with linear dispersion have been extensively studied using Floquet theory \cite{PhysRevB.79.081406,ezawa2013photoinduced,alipourzadeh2023photoinduced}.

Recently, Vargiamidis et al. \cite{vargiamidis2022tunable} investigated the tunable topological phases of monolayer (ML) jacutingaite under magnetic exchange interactions and staggered sublattice potentials. Alipourzadeh et al. \cite{alipourzadeh2023photoinduced,hajati2024electromagnetically} further explored a range of topological phases and spin-valley-polarized currents in ML-jacutingaite subjected to off-resonant optical fields and staggered sublattice potentials. Despite growing interest in Floquet-engineered band structures of ML-jacutingaite, its magneto-optical responses and optoelectronic properties remain largely unexplored.

Given its high sensitivity to external stimuli, including electric, magnetic, and optical fields, comprehensive studies of ML-jacutingaite’s magneto-optical characteristics are essential. Such investigations can uncover critical insights into its underlying quantum behavior and establish its potential for next-generation applications in photonics and optoelectronics.

In this paper, we investigate the spin- and valley-polarized magneto-optical conductivity of monolayer (ML) jacutingaite under the influence of perpendicular magnetic and electric fields, as well as off-resonant circularly polarized laser irradiation, using the Kubo formalism. Our aim is to elucidate the role of Berry curvature in quantum magnetotransport phenomena in jacutingaite materials.
We analyze the energy dispersion of ML-jacutingaite both in the absence and presence of a magnetic field across distinct topological phases. By examining the Berry curvature, we define spin- and valley-polarized Chern numbers as topological invariants. The DOSs are computed for the $K$ and $K'$ valleys in different topological regimes, highlighting the influence of topological phase transitions.
Furthermore, we explore possible magneto-optical transitions between quantized LLs (LLs), governed by optical selection rules. We compute both the real and imaginary components of the magneto-optical conductivity as functions of the external electric and optical fields. Finally, we study the effect of doping, modulated via electron concentration, on the LL structure and its impact on the magneto-optical absorption spectrum.

\begin{figure*}[ht!]
	\centering
	\includegraphics[width=0.8\linewidth]{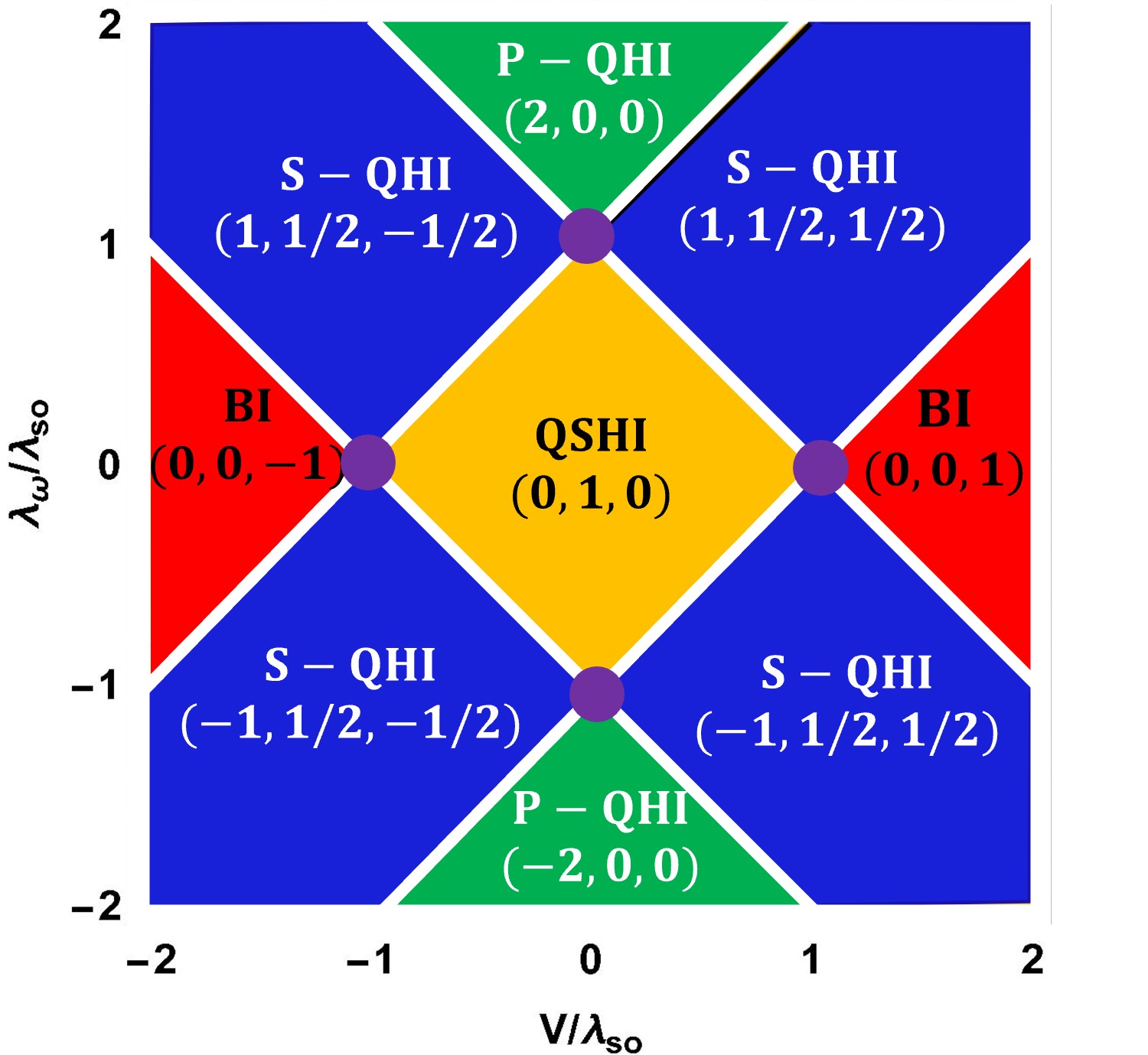}
	\caption{The phase diagram of the monolayer of
jacutingaite as a function of $\Delta_{z}/\Delta_{so}$ and $\lambda_{\omega}/\Delta_{so}$. The distinct electronic phases are labeled by different colors and are indexed by the total, spin, and valley Chern numbers ($\mathcal{C}$, $\mathcal{C}_{s}$ and $\mathcal{C}_{v}$).}
	\label{phase}
\end{figure*}

\begin{figure*}[ht!]
	\centering
	\includegraphics[width=0.450\linewidth]{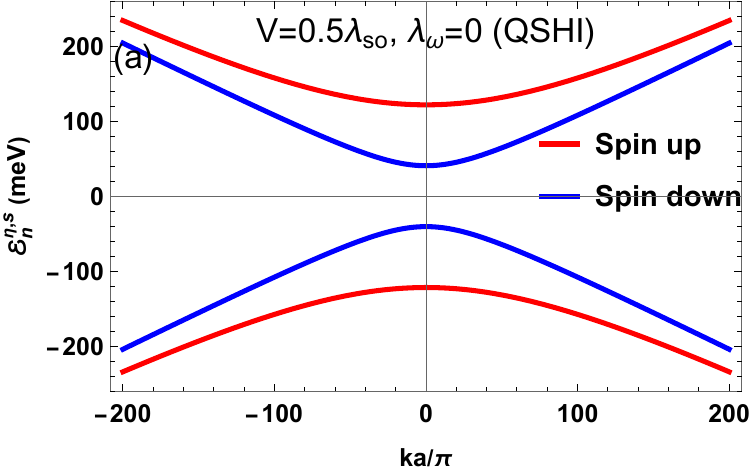}
	\includegraphics[width=0.450\linewidth]{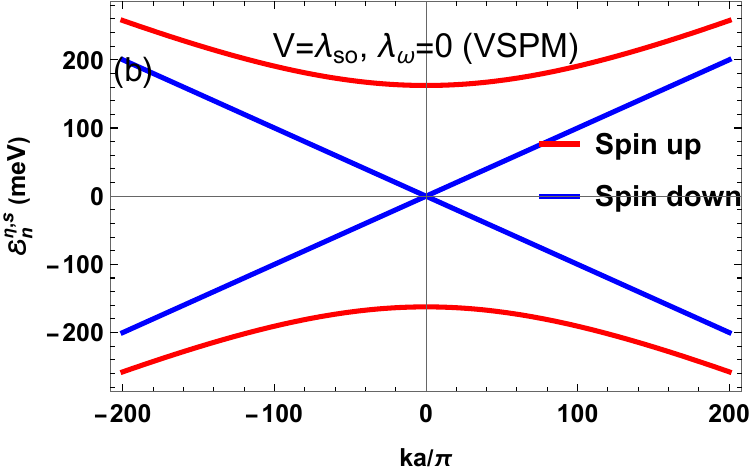}\\
	\includegraphics[width=0.450\linewidth]{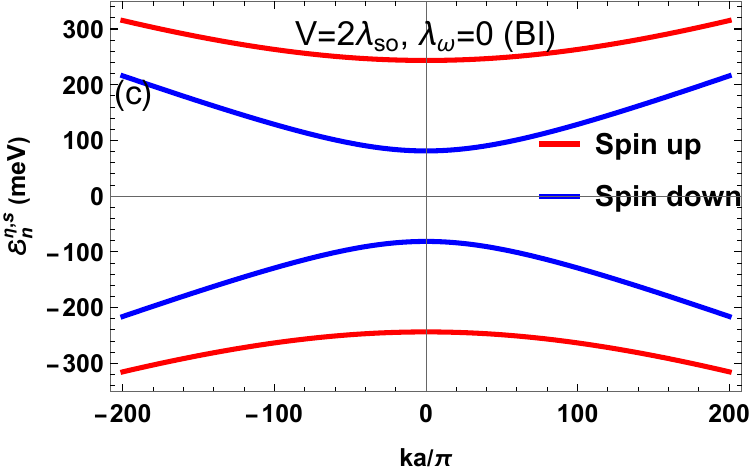}
	\includegraphics[width=0.450\linewidth]{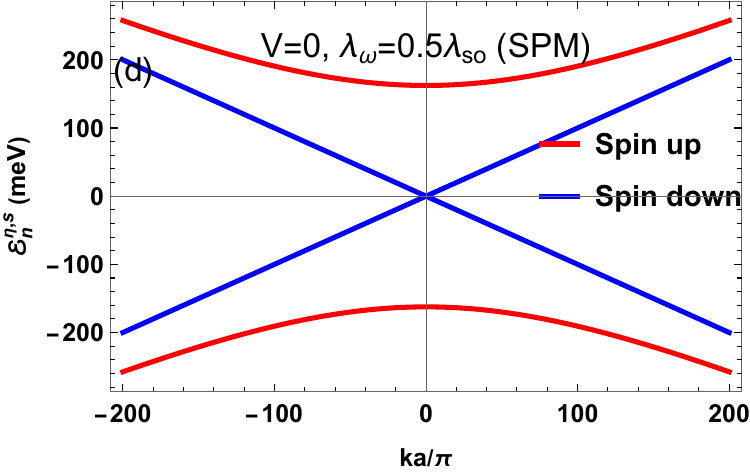}\\
	\includegraphics[width=0.450\linewidth]{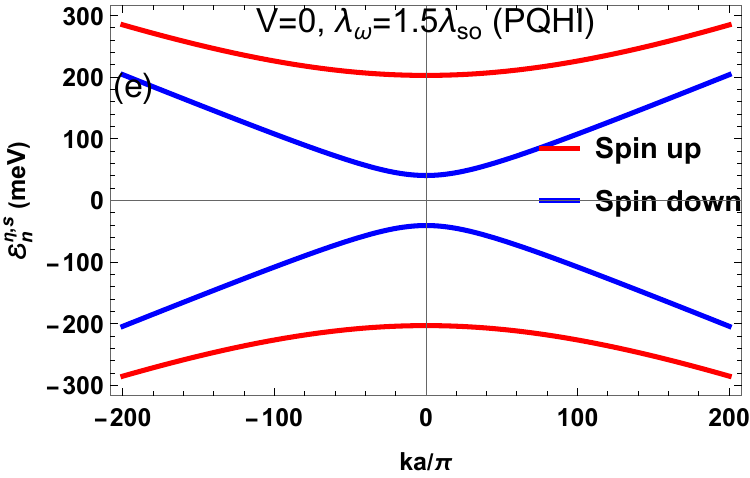}
	\includegraphics[width=0.450\linewidth]{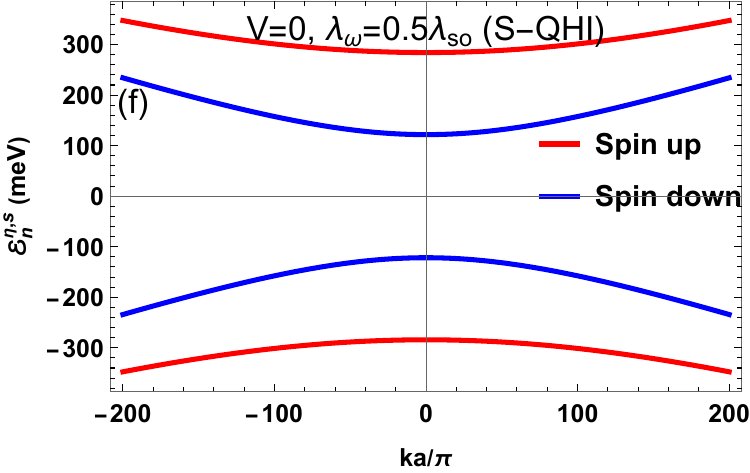}
	\caption{Band structure of jacutingaite in distinct topological phases at the $K$ valley. Q (a) QSHI ($V=0.5\lambda_{so}$, $\lambda_{\omega}=0$), (b) VSPM  ($V=\lambda_{so}$, $\lambda_{\omega}=0$), (c) BI ($V=1.5\lambda_{so}$,  $\lambda_{\omega}=0$), (d) SPM ($V=0$,  $\lambda_{\omega}=\lambda_{so}$), (e) PQHI ($V=0$, $\lambda_{\omega}=1.5\lambda_{so}$) and (f) S-QHI ($V=\lambda_{so}$,  $\lambda_{\omega}=1.5\lambda_{so}$) respectively. The red and blue curves refer to spin-up and spin-down energy bands, respectively.}
	\label{Bandgaps}
\end{figure*}

\subsection{System Hamiltonian and magneto-optical
conductivity}
We consider a pristine monolayer jacutingaite (ML-jacutingaite), subjected to a perpendicular circularly polarized laser field incident on its surface. In the presence of an external out-of-plane electric field, the effective Hamiltonian of ML-jacutingaite at the K and K′ valleys can be expressed as: \cite{ezawa2013photoinduced,vargiamidis2022tunable,alipourzadeh2023photoinduced}
  \begin{equation}\label{a1}
H_{\eta, \mathrm{s}_{\mathrm{z}}}=\hbar v_{\mathrm{f}}\left(\eta k_{\mathrm{x}} \sigma_{\mathrm{x}}+k_{\mathrm{y}} \sigma_{\mathrm{y}}\right)+\eta s \lambda_{so} \sigma_{\mathrm{z}}+\eta \lambda_{\omega} \sigma_{\mathrm{z}}+V \sigma_{\mathrm{z}},
\end{equation}
where, $k_x$ and $k_y$ represent the crystal momentum along $x$ and $y$ directions and the Fermi velocity $v_{F}=3\times10^5$ m/s. The parameter $\eta=\pm 1$ denotes the valley index, distinguishing between the $K$ and $K'$ valleys in momentum space, $\sigma_{i}=(\sigma_{x},\sigma_{y},\sigma_{z})$ are the Pauli matrix in the sublattice space and $s=\pm1$ refers to the electron spin-up and spin-down which is the real spin degrees of freedom. The first term represents the massless graphene-like Hamiltonian, describing low-energy excitations similar to those in graphene. The second term in the Hamiltonian describes the Kane-Mele spin-orbit coupling (SOC) with $\lambda_{so}$= 81.2 meV \cite{marrazzo2018prediction}. Monolayer $\mathrm{Pt_{2}HgSe_{3}}$ is a quantum spin Hall (QSH) insulator with large band gaps and strong SOC, which is nearly 20 times larger—than that in silicene  \cite{ezawa2013spin}. The third term represents the off-resonance circularly polarized optical field, whereas the final term represents the effect of a perpendicular electric field $E_z$, which breaks the inversion symmetry between the $A$ and $B$ sublattices. The energy eigenvalues of Eq. \eqref{a1} can be obtained as
  \begin{equation}\label{b1}
E_{t}^{\eta, s}=t \sqrt{\left(\hbar v_{\mathrm{F}} k\right)^2+\left(\Delta_{\eta, s}\right)^2},
\end{equation}
where $t= \pm$ stands for the electron and hole bands and $\Delta_{\eta, s}=\eta \lambda_{\omega}+\eta s \lambda_{so}+V$ is the Dirac mass  responsible for the effective gap. The off-resonant circularly polarized optical field $\lambda_{\omega}$ and the sublattice staggered potential $V$ break inversion symmetries, leading to nonzero Berry curvature (BC) and, consequently, nonzero anomalous Hall conductivity. The Berry curvature in the out-of-plane
direction for the $t$ band can be obtained as: \cite{shah2022optical}
  \begin{equation}\label{b2}
\Omega_{\eta, \mathrm{s}}^t(k)=-2 \hbar^2 \operatorname{Im} \sum_{\mathrm{t} \neq \mathrm{t}^{\prime}} f_{t}^{\eta, s}(k) \frac{\left\langle\psi_{t}^{\eta, s}\left|v_{\mathrm{x}}\right| \psi_{ t'}^{\eta, s}\right\rangle\left\langle \psi_{ t'}^{\eta, s}\left|v_{\mathrm{y}}\right| \psi_{t}^{\eta, s}\right\rangle}{\left(E_{t}^{\eta, s}-E_{t'}^{\eta, s}\right)^2},
\end{equation}
where $f_{t}^{\eta, s}(k)=\left[1+\exp \left(E_{t}^{\eta, s}-\mu_{\mathrm{f}}\right) / k_{\mathrm{B}} T\right]^{-1}$ is the Fermi Dirac distribution function for the $t$th band with chemical potential $\mu_{F}$, and $\psi_{t}^{\eta, s}$ is the Bloch
state with energy eigenvalues $E_{t}^{\eta, s}$. The interplay between $V$ and $\lambda_{\omega}$ in ML-jacutingaite enables the realization of diverse equilibrium phases, highlighting the system's tunability and rich physics. We determine the various topological states with the help
of the effective Dirac mass $\Delta_{\eta, s}$. In the insulating phase, with the chemical potential within the bulk gap, the spin and valley-dependent Chern number can be expressed as
\begin{equation}\label{b4}
\mathcal{C}_{\eta, s}=\frac{1}{2 \pi} \int d^2 k \Omega_{\eta, s}^t(k).
\end{equation}
The spin and valley polarized Chern number in simplified form can be written as
\begin{equation}\label{b5}
\mathcal{C}_{\eta, \mathrm{s}}=\frac{\eta}{2} \operatorname{sgn}\left(\Delta_{\eta, \mathrm{s}}\right),
\end{equation}
wherein $\operatorname{sgn}(x)$ is the signum function. From Eq. \eqref{b5}, we can characterize the topological insulator state by the charge Chern number and spin Chern number. The spin and valley Chern numbers can be computed as
\begin{equation}\label{b6}
\mathcal{C}_{\mathrm{s}}=\frac{\left(\mathcal{C}_{\uparrow}-\mathcal{C}_{\downarrow}\right)} {2},
\end{equation}
and 
\begin{equation}\label{b7}
\mathcal{C}_{\mathrm{v}}=\frac{\left(\mathcal{C}_{\mathrm{K}}-\mathcal{C}_{\mathrm{K}^{\prime}}\right)}{2},
\end{equation}
where $\mathcal{C}_{\uparrow(\downarrow)}=\sum_\eta C_{\eta, \uparrow(\downarrow)}$ and $\mathcal{C}_{\mathrm{K}\left(\mathrm{K}^{\prime}\right)}=\sum_{\mathrm{s}} \mathcal{C}_{\mathrm{K}, \mathrm{s}\left(\mathrm{~K}^{\prime}, \mathrm{s}\right)}$. The spin and valley Chern number determines the topological properties of the ML-jacutingaite material, with $\mathcal{C}\neq 0$ indicating a nontrivial topological phase. In this case, the sign of the Dirac mass $\Delta_{\eta, s}$ determines the topological phase of the material. A topological phase transition takes place when the sign of the effective
Dirac mass $\Delta_{\eta, s}$
changes. The boundaries between phases are defined by $\Delta_{\eta, s}$= 0 \cite{ezawa2012topological,ezawa2013photoinduced}.

Figure~\ref{phase} summarizes the phase diagram for ML-jacutingaite, showing distinct topological phases separated by phase transition boundaries in the $V/\lambda_{so}$ and $\lambda_{\omega}/\lambda_{so}$ plane. The total, spin, and valley Chern numbers $\left(\mathcal{C}, \mathcal{C}_{\mathrm{s}}\right.$, $\mathcal{C}_{\mathrm{v}}$) serve as topological invariants, distinguishing different phases and labeling them accordingly (represented by distinct colors in phase diagrams). For each topological phase, we present some typical band structures
of ML-jacutingaite in Figs.~\ref{Bandgaps}(a)-(f). The energy spectrum in Figs.~\ref{Bandgaps}(a)-(f) corresponds to different color regions in Fig.~\ref{phase}. External stimuli $V$ and $\lambda_{\omega}$ affect the effective gap and can be used to control the Dirac mass $\Delta_{\eta, s}$ for the different Dirac cones. By modulating the external fields, we can get a wide variety of electronic phases in monolayer jacutingaite. Depending on how many Dirac cones have nonzero mass gaps and non-zero Chern numbers quantum phases, for example, quantum spin Hall insulator (QSHI), spin-polarized
metal (SPM), spin-polarized quantum Hall insulator (S-QHI), spin valley polarized metal (SVPM), normal or band insulator
(BI), single Dirac cone (SDC), and polarized
spin quantum Hall insulator (PQHI) become possible. In the unperturbed jacutingaite system, where both $\lambda_{\omega} = 0$ and $V = 0$, the energy bands are spin-degenerate and separated by an insulating gap of $2\lambda_{\mathrm{so}}$. This corresponds to the QSHI phase, characterized by spin Chern numbers $\mathcal{C}_{\uparrow} = 1$ and $\mathcal{C}_{\downarrow} = -1$, yielding a total spin Chern number $\mathcal{C}_{\mathrm{s}} = 1$. The total Chern number and the valley Chern number remain zero, i.e., $\mathcal{C} = \mathcal{C}_{\mathrm{v}} = 0$, consistent with the findings reported in Ref.~\cite{vargiamidis2022tunable}. Initially, we fix the off-resonant irradiated optical field $\lambda_{\omega}$=0 and tune the staggered sublattice potential $V$.  As long as $V<\lambda_{so}$,
the ML-jacutingaite remains in the QSHI phase with spin splitting and two energy gaps as illustrated in Fig.~\ref{Bandgaps}(a).  However, it should be noted that when $V=\lambda_{so}$, the spin-down effective gap $\left|V-\eta \lambda_{so}\right|$ closes, which reflects the graphene-like nature of the VSPM. The corresponding
band structure of the VSPM state for the $K$ valley is shown in Fig.~\ref{Bandgaps}(b). By further increasing the potential $V$, the system undergoes a quantum phase transition from the QSHI phase to a trivial band insulator (BI) phase. During this transition, the bandgap closes and reopens with a topologically trivial character, as illustrated in Fig.~\ref{Bandgaps}(c). In the BI regime, the spin Chern number $\mathcal{C}_{\mathrm{s}}=0$ while the valley Chern number is $\mathcal{C}_{\mathrm{v}}=1$. The BI regime can be related to the $\mathrm{QVHI}$ state. 
Conversely, if we increase $\lambda_{\omega}$ while keeping $V=0$, the spin-down gap closes at $\lambda_{\omega}=\lambda_{so}$, 
where ML-jacutingaite is a semimetal (Fig.~\ref{Bandgaps}(d)). The spin-polarized metal (SPM) appears at this point. Further increasing $\lambda_{\omega}$ results in the gap reopening, and the system enters a quantum anomalous Hall effect caused by light, namely a P-QHI, as depicted in Fig.~\ref{Bandgaps}(e). The associated Chern numbers to the P-QHI phase are $\mathcal{C}=2$ and $\mathcal{C}_{\mathrm{s}}=\mathcal{C}_{\mathrm{v}}=0$.  What will happen when both the staggered electric potential and off-resonant laser fields are turned on (i.e., $\left| \pm V \pm \lambda_{\omega}\right|>\lambda_{so}$)? When both of the fields are switched on, we can observe
a phase transition from the AQHI to an electromagnetically induced $\mathrm{S}-\mathrm{QHI}$ phase. In the $\mathrm{S}$-$\mathrm{QHI}$ phase, the spin- and valley-resolved Chern numbers are quantized as $\mathcal{C}_{\uparrow} = 1$, $\mathcal{C}_{\downarrow} = 0$, $\mathcal{C}_{\mathrm{K}} = 1$, and $\mathcal{C}_{\mathrm{K}^{\prime}} = 0$. The corresponding band structure of jacutingaite in this regime is displayed in Fig.~\ref{Bandgaps}(f).

\subsection{Monolayer  jacutingaite in a magnetic field}
\begin{figure*}[t!]
	\centering		
	\includegraphics[width=0.4\linewidth]{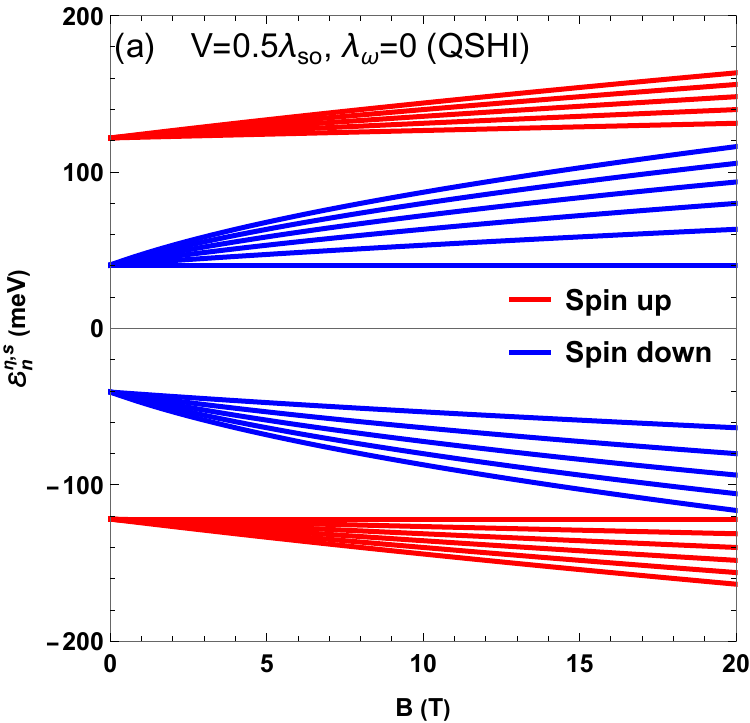}
	\includegraphics[width=0.4\linewidth]{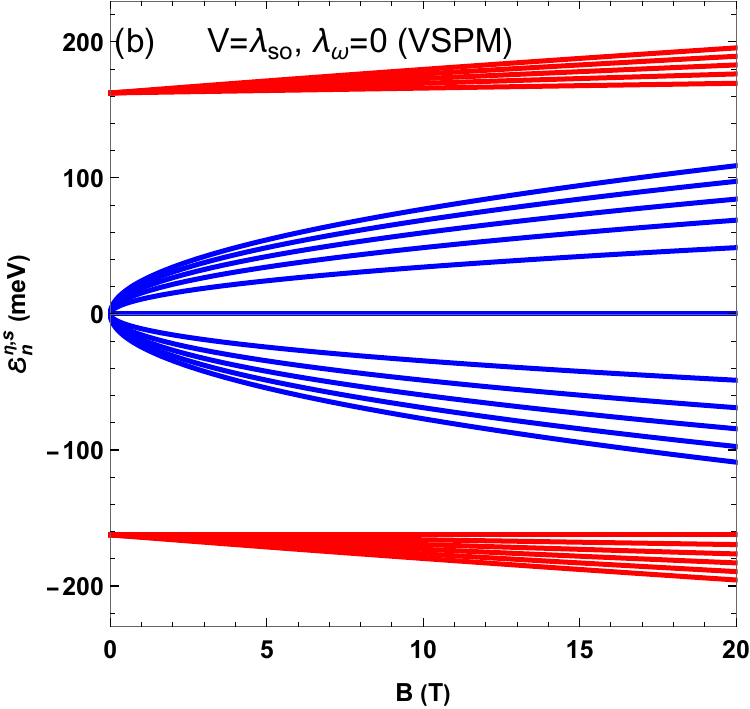}\\
	\includegraphics[width=0.4\linewidth]{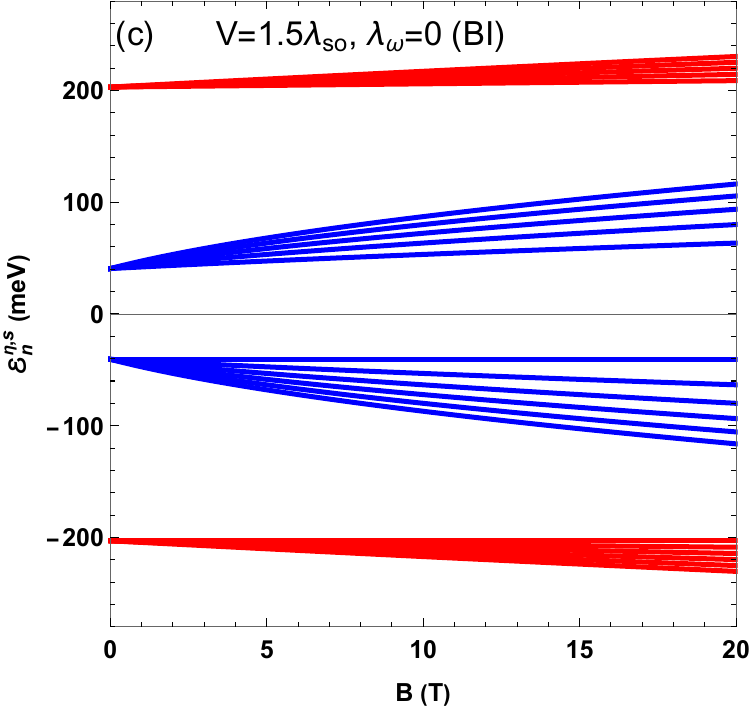}
	\includegraphics[width=0.4\linewidth]{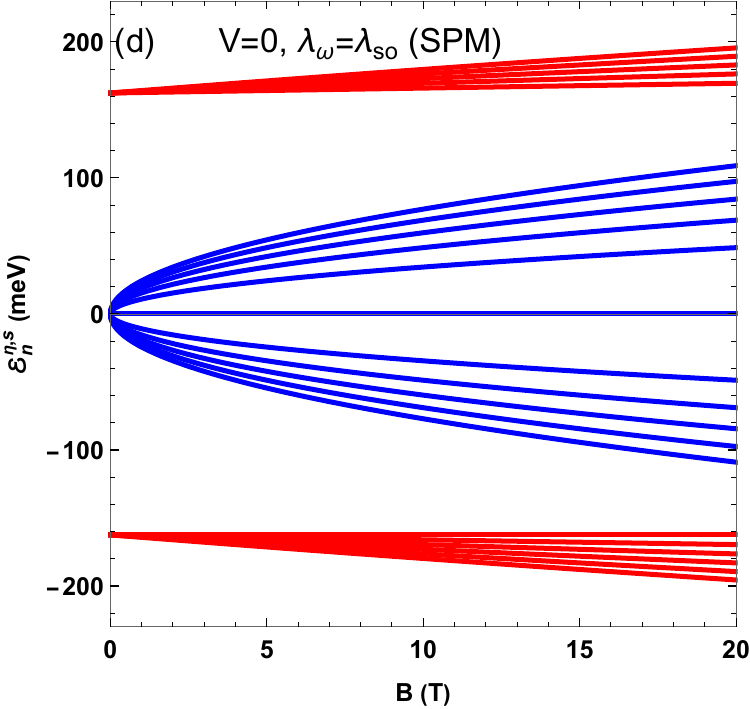}\\
	\includegraphics[width=0.4\linewidth]{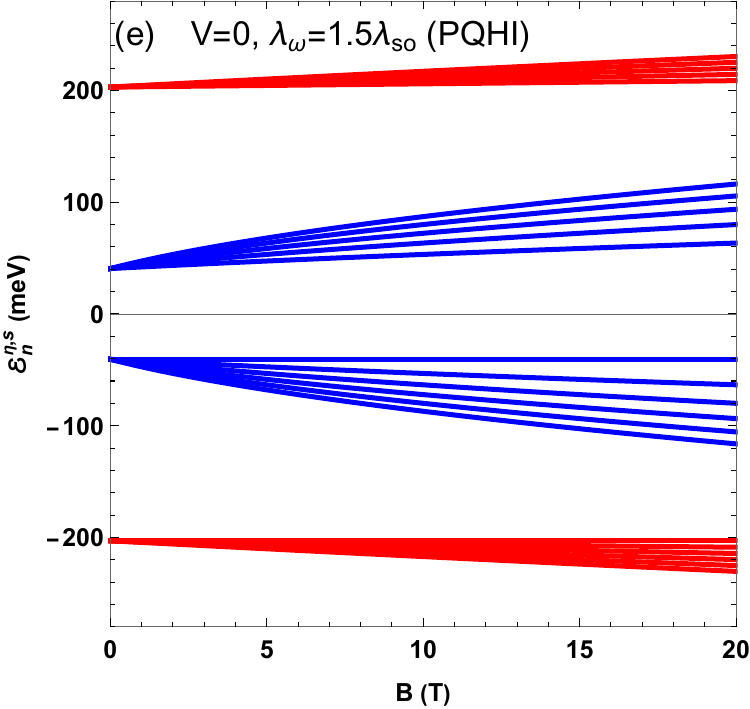}
	\includegraphics[width=0.4\linewidth]{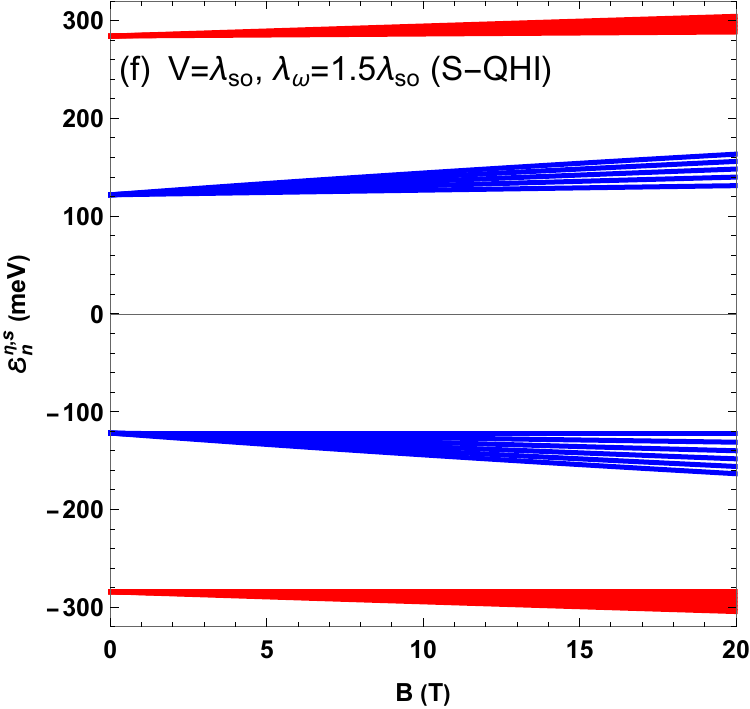}
	\caption{The LLs of
 jacutingaite (in units of meV) as a function of magnetic field (in units of Tesla) strength $B$ in different topological phases. (a) QSHI ($V=0.5\lambda_{so}$ $\lambda_{\omega}=0$), (b) VSPM  ($V=\lambda_{so}$ $\lambda_{\omega}=0$), (c) BI ($V=1.5\lambda_{so}$,  $\lambda_{\omega}=0$), (d) SPM ($V=0$,  $\lambda_{\omega}=\lambda_{so}$), (e) PQHI ($V=0$, $\lambda_{\omega}=1.5\lambda_{so}$) and (f) S-QHI ($V=\lambda_{so}$,  $\lambda_{\omega}=1.5\lambda_{so}$) respectively. The red and blue curves refer to spin-up and spin-down energy bands, respectively. Notice that the LL changes in various phases, for instance, in the QSHI phase the $n=0$ levels are split (one above and one below zero), whereas in the VSPM phase, they meet at zero. Similarly, when the BI phase is reached, the $n=0$ levels invert into the valence band.}
	\label{LL1}
\end{figure*}

A static uniform magnetic field $B$ is applied perpendicular to the $xy$ plane, and the Landau gauge is chosen for the vector potential, $A = (-yB, 0, 0)$. The wave vector alters to ${\bf k}-e{\bf A}$. We diagonalize the Hamiltonian in Eq. \eqref{a1} by modifying the momentum \cite{PhysRevB.81.195431} and obtain the electronic energy eigenvalues as:
	\begin{equation}\label{a2}
	\mathcal{E}^{\eta,s}_{n,t}=\begin{cases}
	t\sqrt{2 v_{F}^{2} \hbar e B |n|+\Delta_{\eta,s}^{2}}, & \text{if $n\neq 0$}.\\
	-\eta \Delta_{\eta,s}, & \text{if $n=0$}.
	\end{cases}
	\end{equation}
Here, $t = \textrm{sgn}(n)$ distinguishes between electron ($t=+1$) and hole ($t=-1$) states, $\Delta_{\eta,s}=\eta s \lambda_{so}+\eta \lambda_{\omega}+V$ and $n$ is an integer quantum number that labels the LLs. The corresponding eigenfunctions at the $K$ and $K'$ valleys are given by
	\begin{align}\label{a3}
	|\psi_{n, t}^{+1, s}\rangle =\begin{pmatrix} -i\mathcal{A}_{n, t}^{+1, s} |\phi_{n-1}\rangle
	\\  \mathcal{B}_{n, t}^{+1, s} |\phi_{n}\rangle \\ \end{pmatrix}
	\end{align}
	and
	\begin{align}\label{a4}
	|\psi_{n, t}^{-1, s}\rangle=\begin{pmatrix} -i\mathcal{A}_{n, t}^{-1, s} |\phi_{n}\rangle
	\\  \mathcal{B}_{n, t}^{-1, s}|\phi_{n-1}\rangle \\ \end{pmatrix},
	\end{align}
where $|\phi_{n}\rangle$ represents an orthonormal Fock state of the quantum harmonic oscillator, corresponding to the $n^\text{th}$ LL wavefunction, and $\mathcal{A}_{n, t}^{\eta, s}$ and $\mathcal{B}_{n, t}^{\eta, s}$ are given by,
	\begin{align}\label{a5}
	\mathcal{A}_{n, t}^{\eta, s} =& \begin{cases}
	\frac{\sqrt{|\mathcal{E}^{\eta,s}_{n,t}|+t\Delta_{\eta,s}}}{\sqrt{2|\mathcal{E}^{\eta,s}_{n,t}|}}, & \text{if $n\neq 0$}.\\
	\frac{1-\eta}{2}, & \text{if $n=0$}.
	\end{cases}
	\end{align}
	and
	\begin{align}\label{a6}
	\mathcal{B}_{n, t}^{\eta, s}= & \begin{cases}
	\frac{\sqrt{|\mathcal{E}^{\eta,s}_{n,t}|-t\Delta_{\eta,s}}}{\sqrt{2|\mathcal{E}^{\eta,s}_{n,t}|}}, & \text{if $n\neq 0$}.\\
	\frac{1+\eta}{2}, & \text{if $n=0$}.
	\end{cases}
	\end{align}
In Figs.~\ref{LL1}(a)-(f), we present the evolution of LLs as a function of the magnetic field $B$ across distinct topological phases at the $K$ valley. Figure~\ref{LL1}(a) displays the LLs of ML-jacutingaite in the QSHI phase as a function of magnetic field. Owing to its stronger SOC, the system exhibits significantly enhanced spin splitting compared to silicene \cite{tabert2013magneto}. The red curves represent spin-up states, while the blue curves denote spin-down states.  In contrast to pristine graphene, the $n=0$ LLs in this system are not pinned at zero energy. These levels are doubly degenerate \cite{PhysRevB.81.195431}, with the spin-up $n=0$ LL appearing at negative energy and the spin-down counterpart at positive energy, as illustrated in Fig.~\ref{LL1}(a). In the QSHI phase, the $n=0$ LL for spin-up electrons is at negative energy, whereas for spin-down electrons, it is at positive energy, as illustrated in Fig.~\ref{LL1}(a). These spin- and valley-split
$n=0$ levels are independent of the magnetic field according to Eq.~\eqref{a2} and can be tuned 
 only through the SOC, off-resonant irradiated optical field $\lambda_{\omega}$, and the staggered sublattice potential $V$. On the other hand, all LLs with $n\neq 0$ in ML-jacutingaite at the $K$ and $K'$ valleys are quadruple spins and valley degenerate and scale as $\sqrt{B}$ similar to graphene. When $\lambda_{\omega}=0$ and $V=\lambda_{so}$, then the $n=0$ spin-down (-up) level at the $K (K')$ point sits at zero energy, which exhibits the graphene-like behavior of the VSPM state. As shown in Fig.~\ref{LL1}(b), 
each LL in the VSPM state is twice spin
degenerate with $K\uparrow = K'\downarrow$ and $K\downarrow = K'\uparrow$. It must be noted that the location of the  $n=0$ LLs in the 
$K$ and $K'$ valleys can be adjusted by tuning the staggered electric and optical fields.

\begin{figure*}[t!]
	\centering		
	\includegraphics[width=0.45\linewidth]{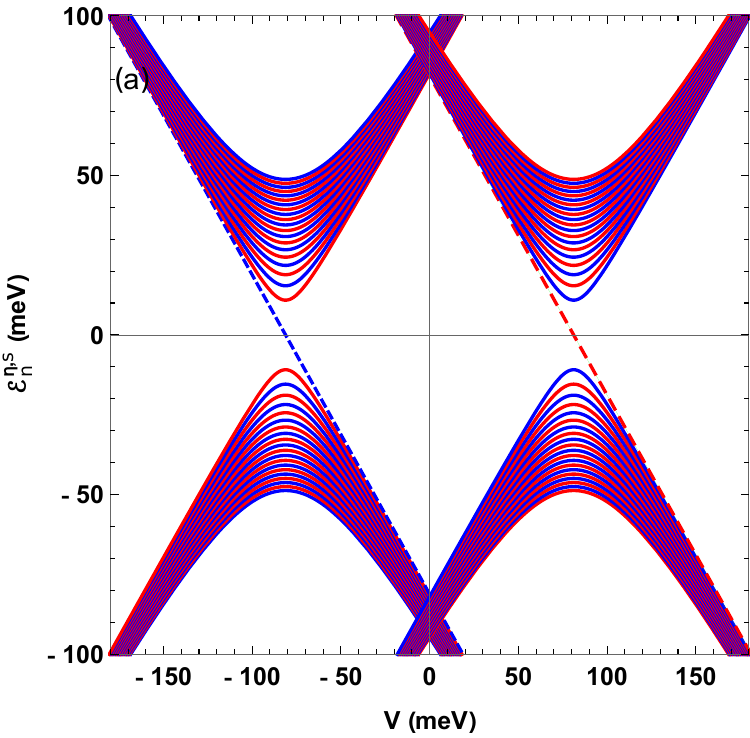}
	\includegraphics[width=0.45\linewidth]{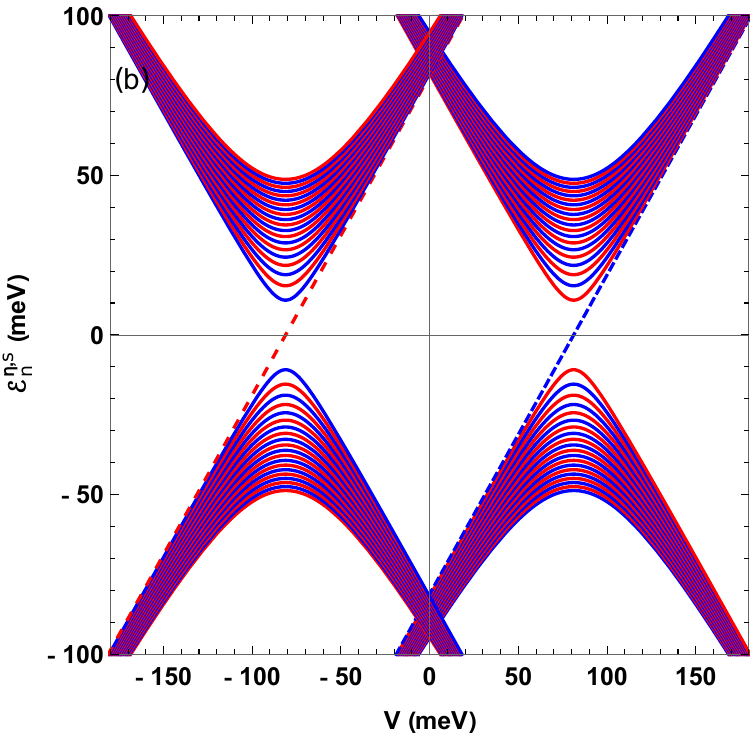}\\
	\includegraphics[width=0.45\linewidth]{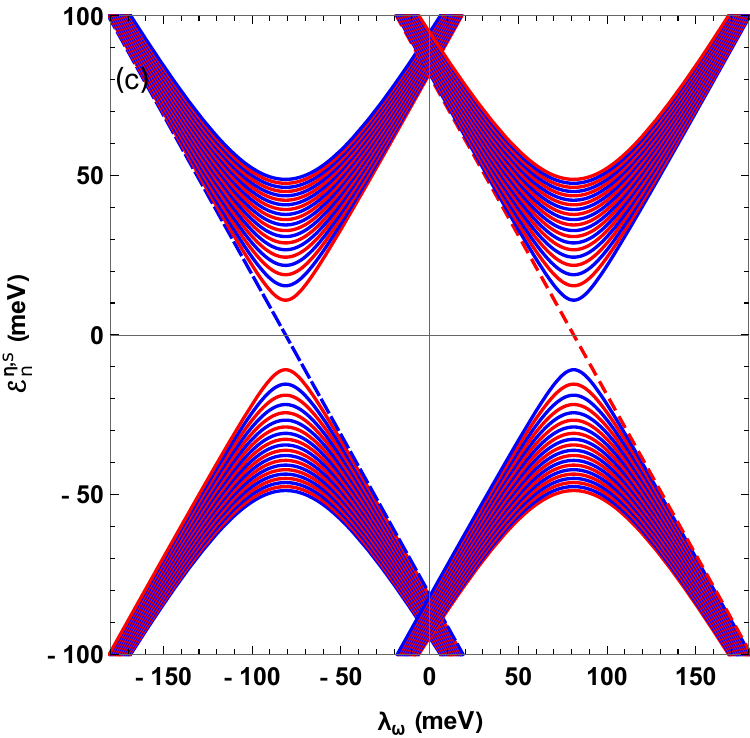}
	\includegraphics[width=0.45\linewidth]{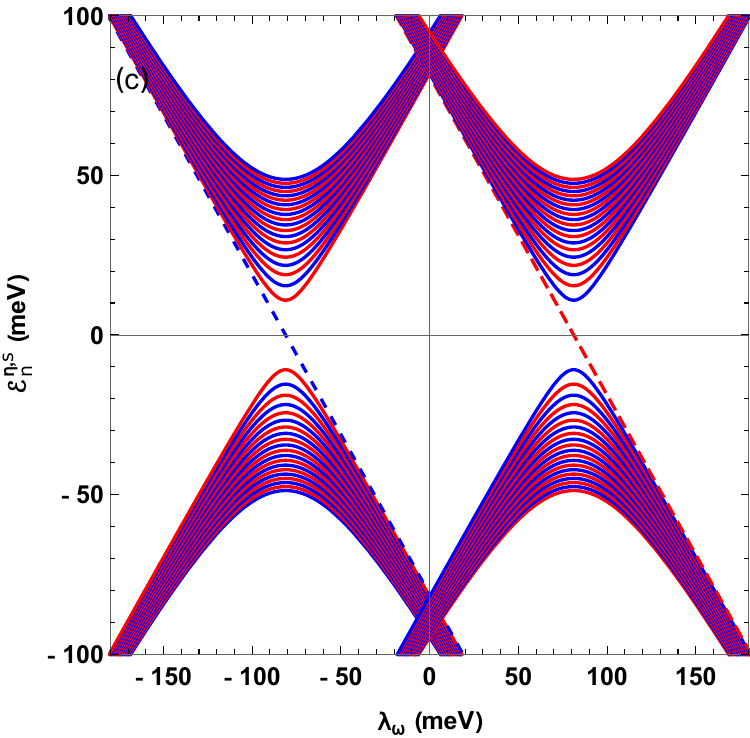}
	\caption{The LLs of
 jacutingaite (in units of meV) as a function of external stimuli (in units of meV) in $K$ and $K'$ valleys. (a)  LL energies as a function of the external electric potential for the fixed optical field in the (a) $K$ valley and (b) $K'$ valley.  LL energies as a function of the external off-resonant irradiated laser field for the fixed staggered electric field in the (c) $K$ valley and (d) $K'$ valley.  The edge states $n=0$ are represented by dashed lines at both valleys: }
	\label{LL2}
\end{figure*}

In Fig.~\ref{LL1}(c), we plot the LLs energy dispersion of the ML-jacutingaite versus the magnetic field at $K$ valley in the BI phase. The zeroth LL plays a crucial role in the transition between the QSHI and BI
phases. In the BI phase, the spin-down energy gap reopens at low energies, and the $n=0$ LLs shift to the valence band, consistent with findings in silicene \cite{shah2021valley} and Floquet
topological insulators thin films \cite{shah2023magneto}. Figure~\ref{LL1}(d) shows the Landau-levels energy dispersion as a function of the magnetic field $B$ for
fixed $V=0$ and finite value of $\lambda_{\omega}=\lambda_{so}$. The SPM state in ML-jacutingaite is similar to the VSPM state, where each LL is twice spin
degenerate as presented in Fig.~\ref{LL1}(d). We have plotted the LLs versus the magnetic field for the PQHI and S-QHI phases in Figs.~\ref{LL1}(e) and (f), respectively. In both phases, the lowest-energy gap of the spin-down band reopens, indicating a significant change in the LL dispersion of the ML-jacutingaite. In the S-QHI state, the gap between the LLs is very large compared to the PQHI phase. 

Figures~\ref{LL2}(a)-(d) show the LLs spectra of ML-jacutingaite as a function of the external stimuli ($V$ and $\lambda_{\mathrm{\omega}}$ in meV) for the $K$ and $K'$ valleys. Again, the red bands
represent spin-up, while the blue bands represent spin-down. LLs for $n= \pm 1$, .... $\pm 20$ in the $K$ valley are shown. We have plotted the LL energy dispersion as a function of staggered electric potential $V$ in Figs.~\ref{LL2}(a) and (b) for the $K$ and $K'$ valleys, respectively. The $n=$0 edge states are denoted by a dashed red (blue) line for spin-up (spin-down).   It should be noted that the edge states switch signs in the $K'$ valley (see  Fig.~\ref{LL2}(b)), which is in agreement with what has been reported in Ref. \cite{calixto2024faraday} for silicene. Similarly, we illustrate the LLs dispersion versus the off-resonant optical field for the $K$ and $K'$ valleys in Figs.~\ref{LL2}(c) and (d), respectively. The edge states originating from the $n=$0 LLs can be seen at the SPM state. Here it can be seen that the edge states do not switch signs in the $K$ and $K'$ valleys, which is reflected in the band structure of ML-jacutingaite \cite{alipourzadeh2023photoinduced}. 

\begin{figure*}[ht!]
	\centering		
	\includegraphics[width=0.45\linewidth]{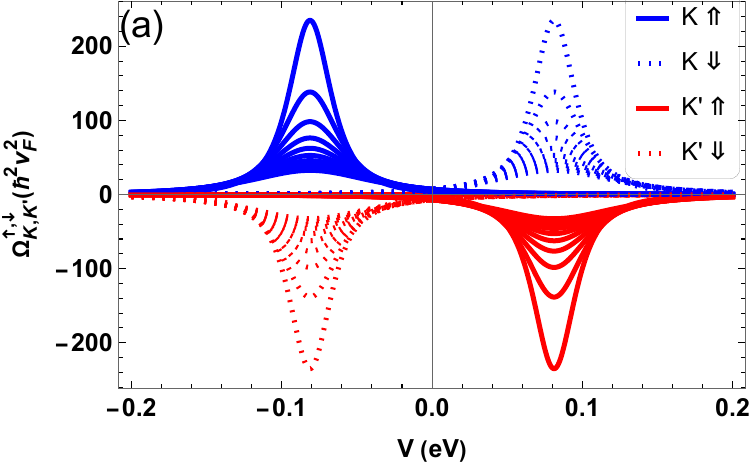}
	\includegraphics[width=0.45\linewidth]{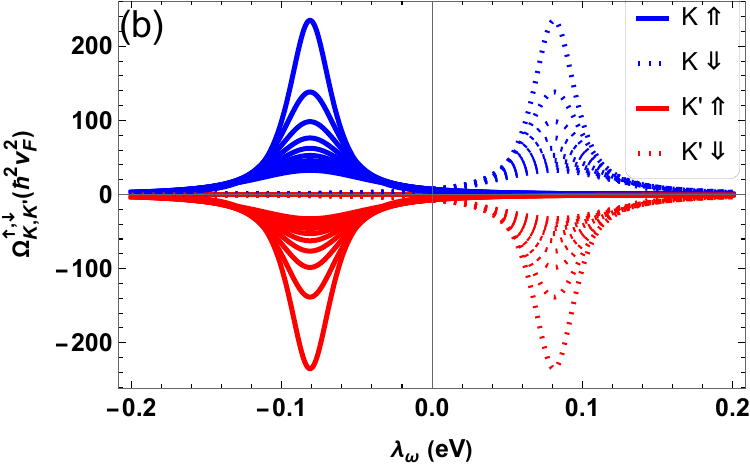}\\
    \includegraphics[width=0.45\linewidth]{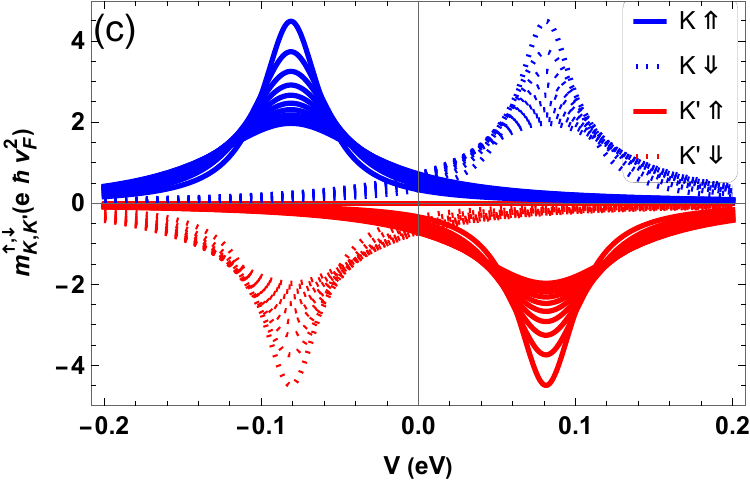}
	\includegraphics[width=0.45\linewidth]{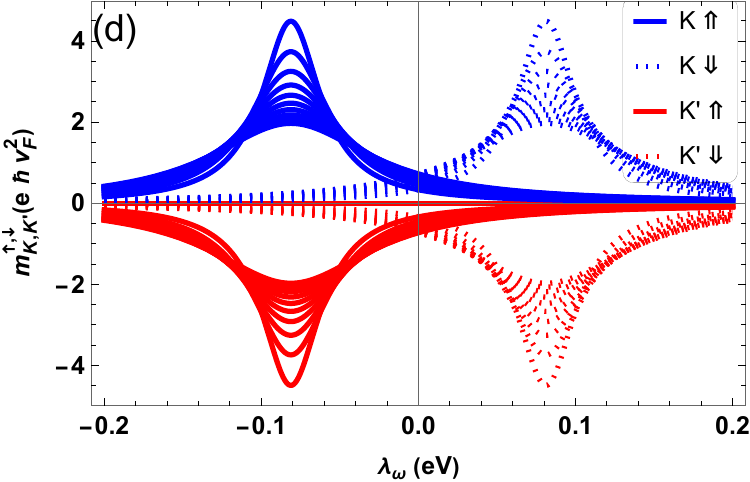}
	\caption{The spin and valley-dependent Berry curvatures of
 jacutingaite as a function of (a) $V$ and (b) $\lambda_{\omega}$ for both spins and valleys. The spin and valley-dependent magnetic moments (in units of $e \hbar v_F^2$) of
 jacutingaite as a function of (c) $V$ and (d) $\lambda_{\omega}$ for both spins and valleys.}
	\label{BC}
\end{figure*}
\subsection{Berry curvature}
Berry curvature is often determined using the Kubo formula, which calculates conductivity based on linear response theory \cite{xiao2010berry}. 
Berry curvature acts like a magnetic field in momentum space, causing electrons to move transversely without an external magnetic field. This leads to a Hall response in each electron state. In systems with LLs above zero, the mathematical properties of Hermite polynomials result in specific behavior due to their orthogonality:
\begin{equation}\label{a72}
\begin{aligned} 
& \Omega^{s}_{\eta}=\frac{\eta \hbar^2 v_{F}^2 \sqrt{n}\hbar \omega}{2 \sqrt{n \hbar^2 w^2+\Delta_{\eta,s}^2}} \\
& \quad \times\left(\frac{1}{\sqrt{n \hbar^2 w^2+\Delta_{\eta,s}^2}+\sqrt{(n+1) \hbar^2 w^2+\Delta_{\eta,s}^2}}\right)^2,
\end{aligned}
\end{equation}
The appearance of $\eta$ in the above expression is very important because it makes the Berry curvature strictly valley-dependent. 
If the inversion and time reversal symmetries are simultaneously preserved, $\Omega_{\eta}$ will vanish. However, it should be noted that the inversion symmetry is broken by the term $\Delta_{\eta,s}$, while the time-reversal symmetry is broken by the presence of a magnetic field.  As seen, the Berry curvature is a spin-valley couple here, and it is significantly larger when $\Delta_{\eta,s}$ becomes small. Importantly, a nonzero Berry curvature implies an anomalous Hall response for each spin/valley sector.

It is clear from the symmetry arguments that:
\begin{equation}\label{a73}
\Omega^{\uparrow}_K=-\Omega^{\downarrow}_{K^{\prime}}, \quad \Omega^{\downarrow}_K=-\Omega^{\uparrow}_{K^{\prime}} \end{equation}

In Figs.~\ref{BC}(a) and (b), we have plotted the Berry curvatures of jacutingaite as functions of electric potential $V$ and optical field $\lambda_{\omega}$, revealing distinct behavior for different spin-valley configurations. As shown in Eq. \eqref{a72}, the Berry curvature is tunable via external fields. We have plotted the Berry curvatures of jacutingaite as a function of $V$ in Fig.~\ref{BC}(a). For negative electric field polarity, the Berry curvature is positive for spin-up electrons in the $K$ valley and negative for spin-down electrons in the $K'$ valley, as illustrated in Fig.~\ref{BC}(a). In contrast, for positive electric field polarity, the Berry curvature is positive for spin-down electrons in the $K$ valley and negative for spin-up electrons in the $K'$ valley. The Berry curvature of jacutingaite as a function of $\lambda_{\omega}$ is shown in Fig.~\ref{BC}(b).  The curvature exhibits opposite signs for spin-up electrons in the $K$ and $K'$ valleys, depending on the optical field polarity: positive curvature in $K$ ($K'$) valley for negative (positive) optical field, and vice versa.

The magnetic moment, which arises from the self-rotation of the wave packet in a valley and contributes to valley-dependent magnetization, in ML- jacutingaite is given as \cite{PhysRevB.107.235417}:
\begin{equation}\label{a76}
\begin{aligned} 
& m^{s}_{\eta}=\frac{e}{\hbar}\Omega^{s}_{\eta}\mathcal{E}^{\eta,s}_{n,t}
\end{aligned}
\end{equation}
In Figs.~\ref{BC}(c) and (d), we have displayed the spin and valley-dependent magnetic moments of
jacutingaite as a function of $V$ and $\lambda_{\omega}$ respectively.

\subsection{DOSs}
Now we want to calculate the electronic DOSs (DOS) of ML-jacutingaite at $K$ and $K'$ valleys. The number of electronic states at a specific energy can be written as \cite{tabert2013magneto}
\begin{equation}\label{a7}
\mathcal{N}(\omega)=\sum_{\eta, s} \sum_k \delta[\omega-\mathcal{E}^{\eta,s}_{n,t}(\boldsymbol{k})].
\end{equation}
The magnetic field $B$ breaks time-reversal symmetry, making the wave vector $k$ no longer a good quantum number. The electronic DOS can be written as
\begin{equation}\label{a8}
\mathcal{N}(\omega)=\mathcal{G} \sum_{\eta, \sigma} \sum_{\substack{N=0 \\ s= \pm}}^{\infty} \delta\left(\omega-\mathcal{E}^{\eta,s}_{n,t}\right),
\end{equation}
where the spin and valley degeneracy of each LL is given by $\mathcal{G}=e B /(2 \pi \hbar c)$, where $c=1$ is taken for simplicity. The low-energy DOSs for ML-jacutingaite at $K$ and $K'$ valleys in the presence of a magnetic field is
\begin{equation}\label{a9}
\mathcal{N}(\omega)=\frac{e B}{2 \pi \hbar} \sum_{\eta, s= \pm} \sum_{\substack{n=0 \\ t= \pm}}^{\infty} \delta\left(\omega-\mathcal{E}^{\eta,s}_{n,t}\right),
\end{equation}
where the $\delta$-function can be evaluated numerically by using the  Lorentzian representation.

\begin{figure*}[ht!]
	\centering		
	\includegraphics[width=0.45\linewidth]{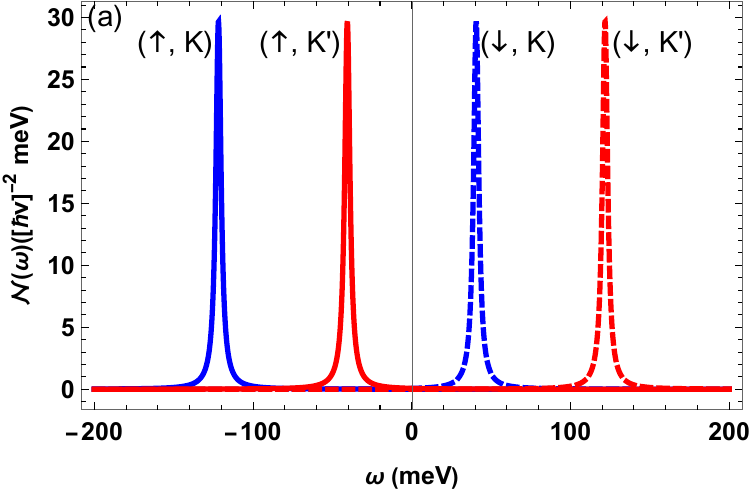}
	\includegraphics[width=0.45\linewidth]{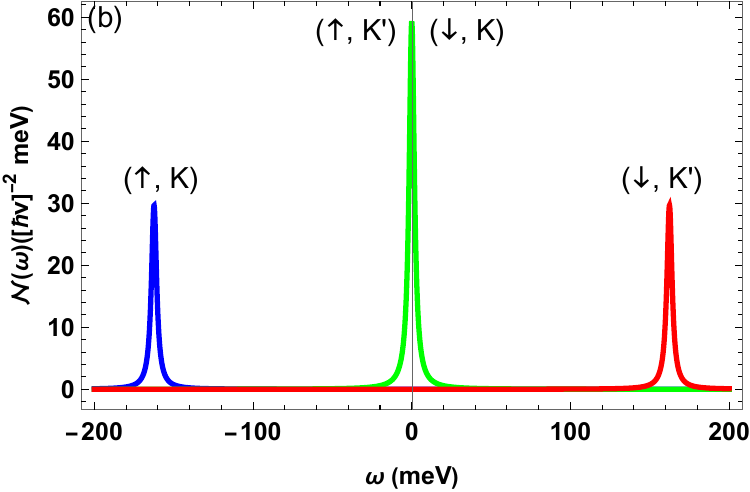}\\
	\includegraphics[width=0.45\linewidth]{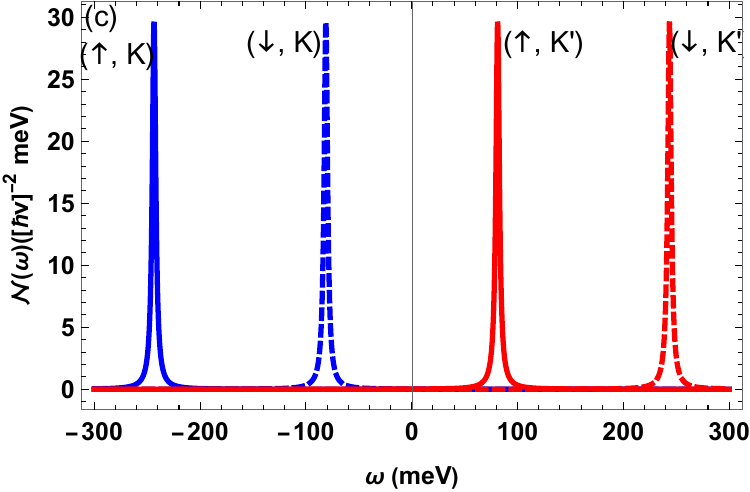}
	\includegraphics[width=0.45\linewidth]{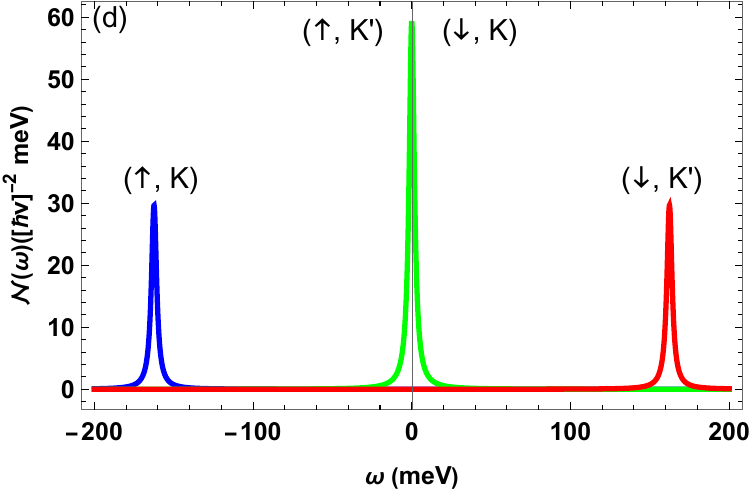}\\
	\includegraphics[width=0.45\linewidth]{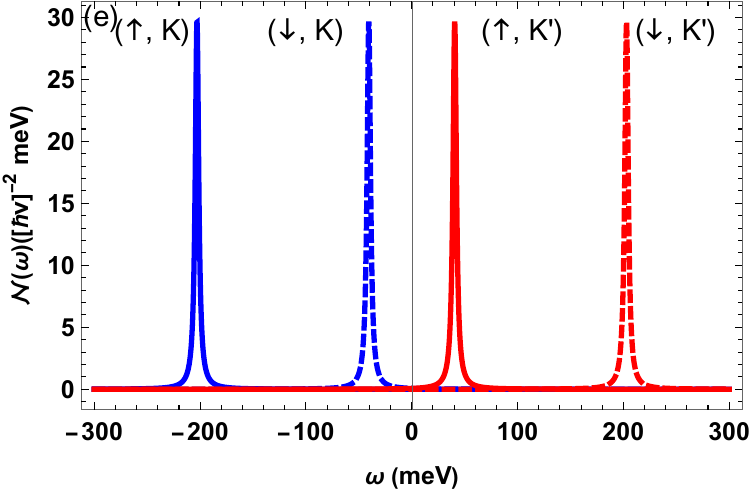}
	\includegraphics[width=0.45\linewidth]{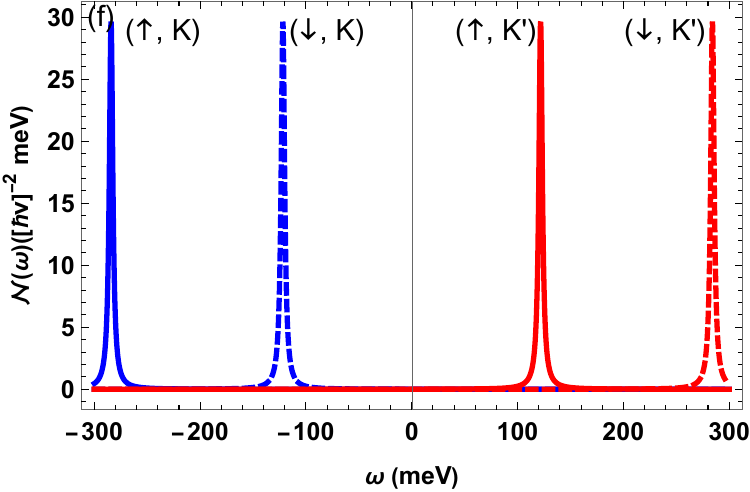}
	\caption{The spin and valley-dependent DOSs of
 jacutingaite as a function of $\hbar \omega$ (in units of meV) (PLEASE CHANGE $\omega$ TO $E$ avoiding the confusion with photon energy)in different topological phases. (a) QSHI ($V=0.5\lambda_{so}$ $\lambda_{\omega}=0$), (b) VSPM  ($V=\lambda_{so}$ $\lambda_{\omega}=0$), (c) BI ($V=1.5\lambda_{so}$,  $\lambda_{\omega}=0$), (d) SPM ($V=0$,  $\lambda_{\omega}=\lambda_{so}$), (e) PQHI ($V=0$, $\lambda_{\omega}=1.5\lambda_{so}$) and (f) S-QHI ($V=\lambda_{so}$,  $\lambda_{\omega}=1.5\lambda_{so}$) respectively.}
	\label{DOS}
\end{figure*}

Figures~\ref{DOS}(a)-(f) present the electronic DOS for ML-jacutingaite at the $K$ and $K'$ valleys across different topological phases. In the QSHI phase [Fig.~\ref{DOS}(a)], four spin- and valley-polarized levels emerge at the magneto-excitation energies corresponding to the $n=0$ LLs. The spin-up (spin-down) states at the $K$ valley are represented by solid (dashed) blue curves, while the spin-down (spin-up) states at the $K'$ valley are depicted by solid (dashed) red curves. Figure~\ref{DOS}(b) displays the spin- and valley-resolved DOS in the VSPM phase as a function of $\omega$ for both valleys. In the VSPM state, the two middle $n=0$ LL peaks merge, forming a large valley-spin degenerate peak (green curve). In the BI phase, the spin-up and spin-down electrons at the $K$ valley reside at negative magneto-excitation energies, while those at the $K'$ valley lie at positive energies as depicted in Fig.~\ref{DOS}(c). The SPM phase exhibits spin- and valley-polarized peaks, similar to the VSPM phase but with distinct ordering as can be seen in Fig.~\ref{DOS}(d). Figures~\ref{DOS}(e) and (f) show the spin-valley-resolved DOS in the PQHI and S-QHI phases, respectively. In the PQHI phase, the energy splitting between the spin-up and spin-down peaks (from $n=0$ LLs) shrinks for both valleys. Conversely, in the S-QHI phase, this splitting widens, reflecting enhanced spin-valley polarization. 

\subsection{Magneto-optical conductivity for the ML-jacutingaite}
The magneto-optical conductivity of monolayer jacutingaite is calculated analytically via the Kubo formula, utilizing the system's eigenvalues and eigenfunctions under external fields. The analytical expressions for the conductivity can be written as~\cite{tse2011magneto},
\begin{widetext}
\begin{equation}\label{a10}
	\sigma_{\mu\nu}(\omega)
	=\frac{i\hbar}{2\pi l_{B}^{2}}\sum_{s, \eta=\pm 1}\sum_{mn}\frac{f_{n, t}^{\eta, s}-f_{m, t'}^{\eta, s}}{E_{n, t}^{\eta, s}-E_{m, t'}^{\eta, s}}\thickspace
	\frac{\langle \psi_{n, t}^{\eta, s}|\hat{j}_{\mu}|\psi_{m, t'}^{\eta, s}\rangle\langle \psi_{m, t'}^{\eta, s}|\hat{j}_{\nu}|\psi_{n, t}^{\eta, s}\rangle}{\hbar \omega-(E_{n, t}^{\eta, s}-E_{m, t'}^{\eta, s})+i\Gamma},
	\end{equation}
where the Fermi--Dirac distribution function is given by
\begin{equation}
f_{n, t}^{\eta, s} = \frac{1}{1 + \exp\left(\frac{E_{n, t}^{\eta, s} - \mu_F}{k_B T}\right)},
\end{equation}
which describes the probability of occupation of a state with energy $E_{n, t}^{\eta, s}$ at temperature $T$ and chemical potential $\mu_F$. The current operator is defined as \( \hat{j}_\mu = e v_F \hat{s}_\mu \), where \( \hat{s}_\mu \) are the Pauli matrices, and the index \( \mu \in \{x, y, z\} \) denotes the spatial direction.  \( E_n \) denotes the energy of the \( n \)-th LL, \( \Gamma \) is the transport scattering rate accounting for the broadening of energy levels, and \( l_B = \sqrt{\hbar / eB} \) is the magnetic length.
We now proceed to evaluate the matrix elements, beginning with the longitudinal conductivity (i.e., for \( \mu = \nu = x \)). The relevant matrix element of the current operator is
\begin{equation}
\left\langle \psi_{n, t}^{\eta, s} \middle| \hat{j}_x \middle| \psi_{m, t'}^{\eta, s} \right\rangle 
= e v \eta \left\langle \psi_{n, t'}^{\eta, s} \middle| \sigma_x \middle| \psi_{m, t}^{\eta, s} \right\rangle,
\end{equation}
where \( \sigma_x \) is the standard pseudospin Pauli matrix. We have,
\begin{equation}\label{a11}
\begin{aligned}
\left\langle\psi_{m, t'}^{\eta, s}\left|\sigma_{x}\right| \psi_{n, t}^{\eta, s}\right\rangle & =\left(\begin{array}{ll}
i \mathcal{A}_{m, t'}^{\eta, s}\langle \phi_{m-1}| & \mathcal{B}_{m, t'}^{\eta, s}\langle \phi_{m}|
\end{array}\right)\left(\begin{array}{cc}
0 & 1 \\
1 & 0
\end{array}\right)\binom{-i \mathcal{A}_{n, t}^{\eta, s}|\phi_{n-1}\rangle}{\mathcal{B}_{n, t}^{\eta, s}|\phi_{n}\rangle}, \\
& =i \mathcal{A}_{m, t'}^{\eta, s} \mathcal{B}_{n, t}^{\eta, s} \delta_{m-1, n}-i \mathcal{B}_{m, t'}^{\eta, s} \mathcal{A}_{n, t}^{\eta, s} \delta_{m+1, n},
\end{aligned}
\end{equation}
and
\begin{equation}\label{a12}
\left\langle\psi_{n, t}^{\eta, s}\left|\sigma_{x}\right| \psi_{m, t'}^{\eta, s}\right\rangle=i \mathcal{A}_{n, t}^{\eta, s} \mathcal{B}_{m, t'}^{\eta, s} \delta_{m n-1}-i \mathcal{B}_{n, t}^{\eta, s} \mathcal{A}_{m, t'}^{\eta, s} \delta_{m-1, n}.
\end{equation}.

Therefore,
\begin{equation}\label{a13}
\begin{aligned}
\left\langle\psi_{n, t}^{\eta, s}\left|\sigma_{x}\right| \psi_{m, t'}^{\eta, s}\right\rangle\left\langle\psi_{m, t'}^{\eta, s}\left|\sigma_{x}\right| \psi_{n, t}^{\eta, s}\right\rangle=\left(\mathcal{A}_{m, t'}^{\eta, s} \mathcal{B}_{n, t}^{\eta, s}\right)^2 & \delta_{m-1, n}+\left(\mathcal{B}_{m, t'}^{\eta, s} \mathcal{A}_{n, t}^{\eta, s}\right)^2 \delta_{m+1, n} \\
& -2 \mathcal{A}_{m, t'}^{\eta, s} \mathcal{B}_{n, t}^{\eta, s} \mathcal{B}_{m, t'}^{\eta, s} \mathcal{A}_{n, t}^{\eta, s} \delta_{m-1, n} \delta_{m+1, n}.
\end{aligned}
\end{equation}
Note that the last term is never nonzero; thus,
\begin{equation}\label{a14}
\left\langle\psi_{n, t}^{\eta, s}\left|\sigma_{x}\right| \psi_{m, t'}^{\eta, s}\right\rangle\left\langle\psi_{m, t'}^{\eta, s}\left|\sigma_{x}\right| \psi_{n, t}^{\eta, s}\right\rangle=\left(\mathcal{A}_{m, t'}^{\eta, s} \mathcal{B}_{n, t}^{\eta, s}\right)^2 \delta_{m-1, n}+\left(\mathcal{B}_{m, t'}^{\eta, s} \mathcal{A}_{n, t}^{\eta, s}\right)^2 \delta_{m+1, n}.
\end{equation}
The Kronecker \( \delta \)-functions indicate that optical transitions are only allowed between states with indices \( m \) and \( n \pm 1 \), in accordance with the harmonic oscillator selection rules for LL transitions. The longitudinal magneto-optical conductivity can be obtained as \cite{tabert2013magneto,shah2019magneto}
\begin{equation}\label{a15}
\begin{aligned}
\sigma_{x x}(\omega)=\frac{i \hbar e^2 v^2}{2 \pi l_B^2} \sum_{\eta, s= \pm} \sum_{n, m=0}^{\infty} \sum_{t, t^{\prime}= \pm} & \frac{\Theta\left(\mu_{F}-E_{m, t'}^{\eta, s}\right)-\Theta\left(\mu_{F}-E_{n, t}^{\eta, s}\right)}{E_{n, t}^{\eta, s}-E_{m, t'}^{\eta, s}} \\
& \times \frac{\left(\mathcal{A}_{m, t'}^{\eta, s} \mathcal{B}_{n, t}^{\eta, s}\right)^2 \delta_{m-1, n}+\left(\mathcal{B}_{m, t'}^{\eta, s} \mathcal{A}_{n, t}^{\eta, s}\right)^2 \delta_{m+1, n}}{\hbar \omega+E_{m, t'}^{\eta, s}-E_{n, t}^{\eta, s}+i \hbar /(2 \tau)} .
\end{aligned}
\end{equation}
The real and imaginary components can be expressed as 
\begin{equation}\label{a16}
\begin{aligned}
\operatorname{Re} & \bigg(\frac{\sigma_{x x}(\omega)}{\sigma_0}\bigg)=\frac{2 v^2 \hbar e B}{\pi} \sum_{\eta, s= \pm} \sum_{n, m=0}^{\infty} \sum_{t, t^{\prime}= \pm} \frac{\Theta\left(\mu_{F}-E_{m, t'}^{\eta, s}\right)-\Theta\left(\mu_{F}-E_{n, t}^{\eta, s}\right)}{E_{n, t}^{\eta, s}-E_{m, t'}^{\eta, s}} \\
& \times\left[\left(\mathcal{A}_{m, t'}^{\eta, s} \mathcal{B}_{n, t}^{\eta, s}\right)^2 \delta_{n, m-\eta}+\left(\mathcal{B}_{m, t'}^{\eta, s} \mathcal{A}_{n, t}^{\eta, s}\right)^2 \delta_{n, m+\eta}\right] \frac{\Gamma}{\Gamma^2+\left(\hbar \omega+E_{m, t'}^{\eta, s}-E_{n, t}^{\eta, s}\right)^2},
\end{aligned}
\end{equation}
and 
\begin{equation}\label{a17}
\begin{aligned}
\operatorname{Im} & \bigg(\frac{\sigma_{x x}(\omega)}{\sigma_0}\bigg)=\frac{2 v^2 \hbar e B}{\pi} \sum_{\eta, \sigma= \pm} \sum_{n, m=0}^{\infty} \sum_{t, t^{\prime}= \pm} \frac{\Theta\left(\mu_{F}-E_{m, t'}^{\eta, s}\right)-\Theta\left(\mu_{F}-E_{n, t}^{\eta, s}\right)}{E_{n, t}^{\eta, s}-E_{m, t'}^{\eta, s}} \\
& \times\left[\left(\mathcal{A}_{m, t'}^{\eta, s} \mathcal{B}_{n, t}^{\eta, s}\right)^2 \delta_{n, m-\eta}+\left(\mathcal{B}_{m, t'}^{\eta, s} \mathcal{A}_{n, t}^{\eta, s}\right)^2 \delta_{n, m+\eta}\right] \frac{\hbar \omega+E_{m, t'}^{\eta, s}-E_{n, t}^{\eta, s}}{\Gamma^2+\left(\hbar \omega+E_{m, t'}^{\eta, s}-E_{n, t}^{\eta, s}\right)^2},
\end{aligned}
\end{equation}
where $\sigma_0=e^2 /(4 \hbar)$ represents the universal conductivity of graphene.
  \begin{figure*}[t!]
	\centering		
	\includegraphics[width=0.450\linewidth]{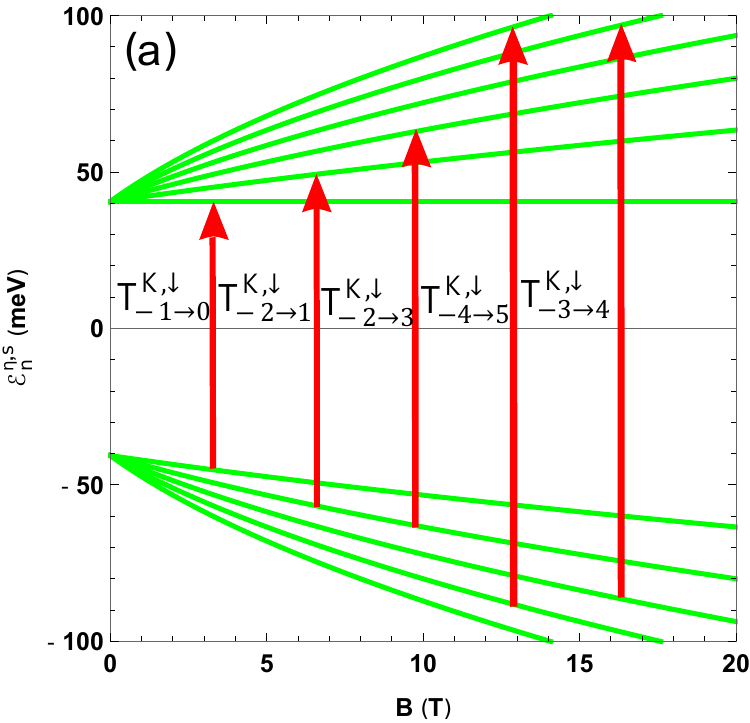}
 \includegraphics[width=0.450\linewidth]{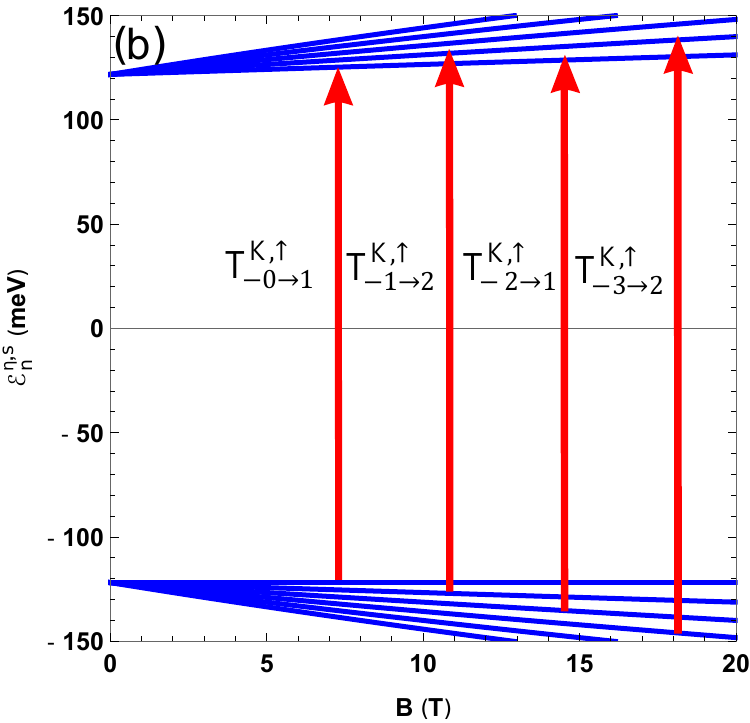}
	\caption{A sketch of LLs (in units of meV) in  ML-jacutingaite as a function of $B$ (in units of Tesla) field with allowed LL transitions in the QSHI regime at the $K$ valley.  LLs and magneto-optical transitions between different LLs for (a) spin-down and (b) spin-up electrons.}
	\label{LL1_graphene}
\end{figure*}

\begin{figure*}[ht!]
	\centering		
	\includegraphics[width=0.450\linewidth]{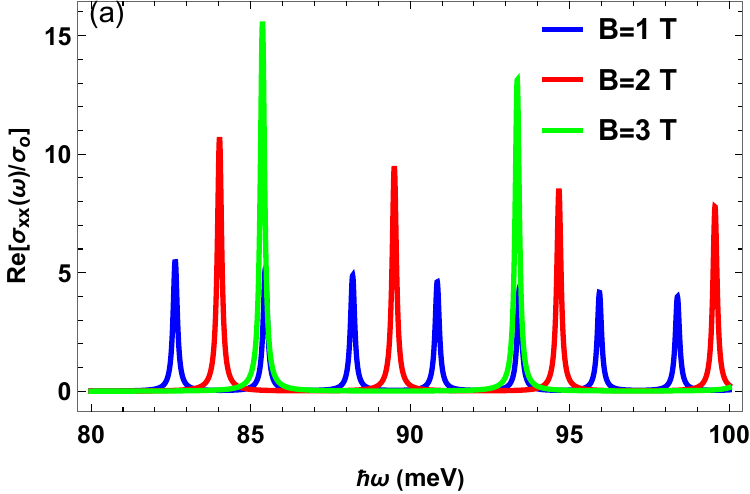}
	\includegraphics[width=0.450\linewidth]{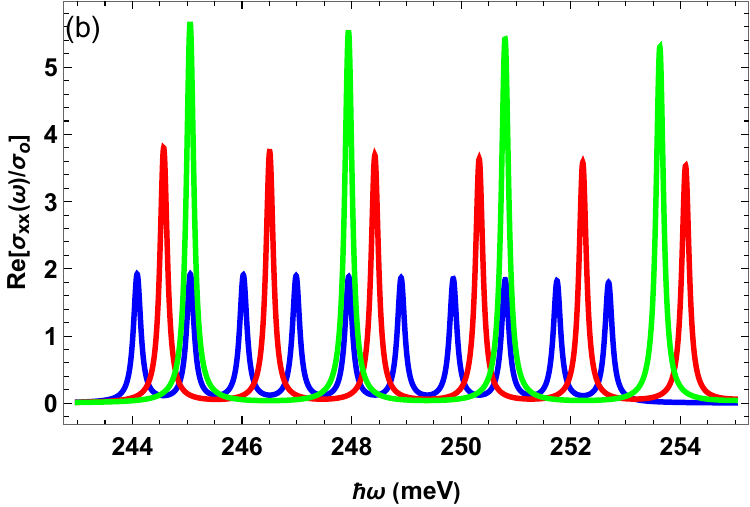}\\
	\includegraphics[width=0.450\linewidth]{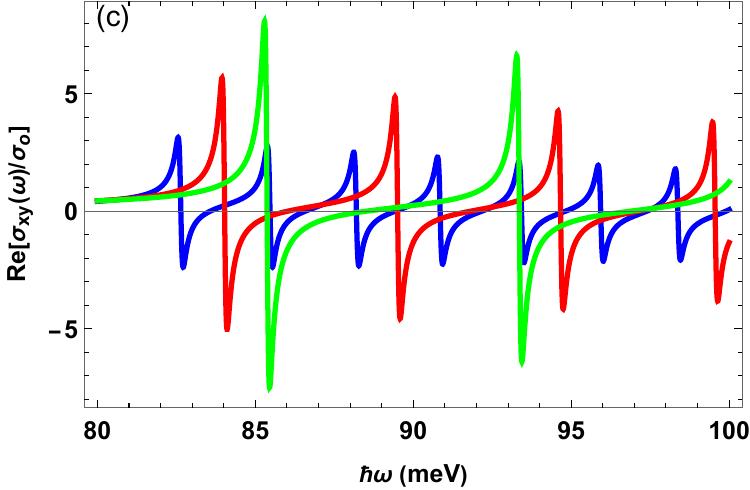}
	\includegraphics[width=0.450\linewidth]{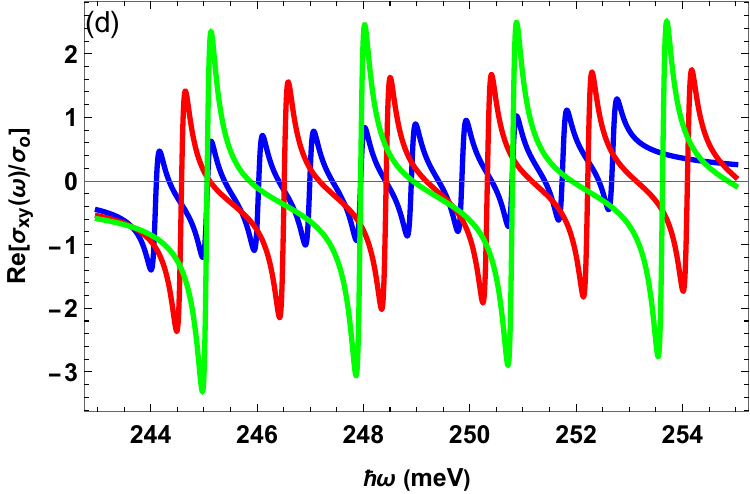}
	\caption{The real part of $\sigma_{xx}$ as a function of incident photon energy $\hbar\omega$ for three different values of the magnetic field strength and a fixed chemical potential $\mu_{F}=0$ in the QSHI phase for the (a) spin-up and (b) spin-down electrons in the $K$ valley.  The real part of $\sigma_{xy}$ as a function of photon energy for three different values of the magnetic field strength and a fixed chemical potential $\mu_{F}=0$ in the QSHI phase for the (c) spin-up and (d) spin-down electrons in the $K$ valley.}
	\label{Sigma1}
\end{figure*}
\begin{figure*}[ht!]
	\centering		
	\includegraphics[width=0.450\linewidth]{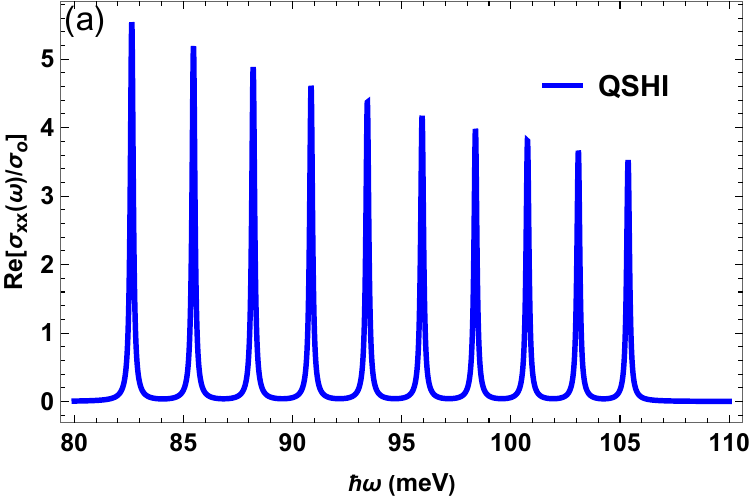}
	\includegraphics[width=0.450\linewidth]{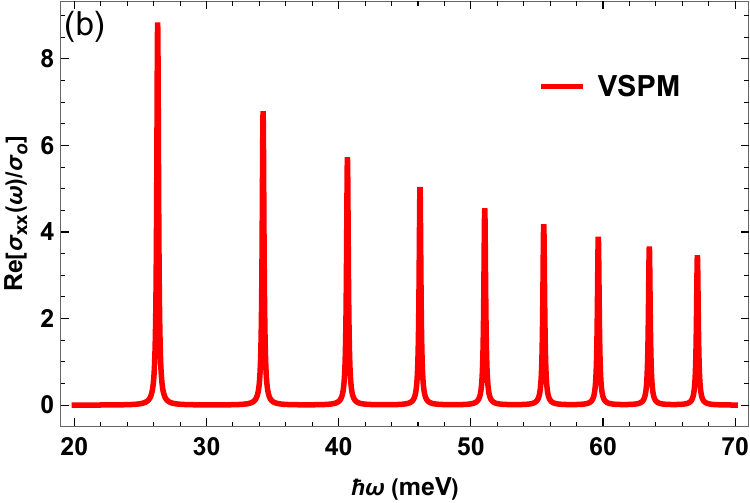}\\
	\includegraphics[width=0.450\linewidth]{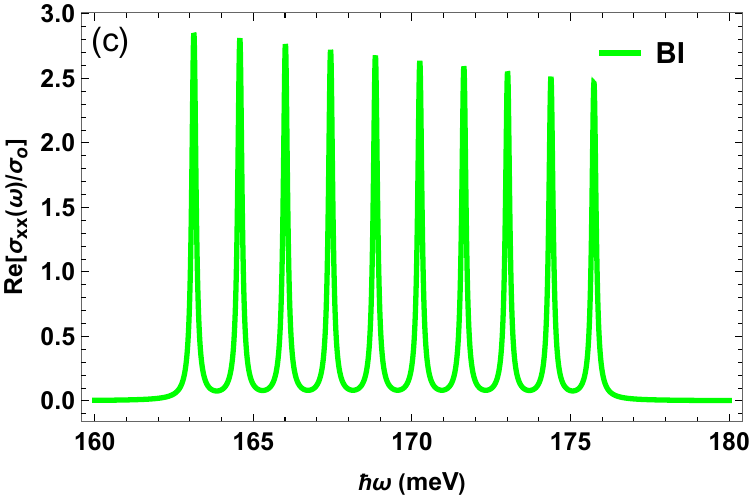}
	\includegraphics[width=0.450\linewidth]{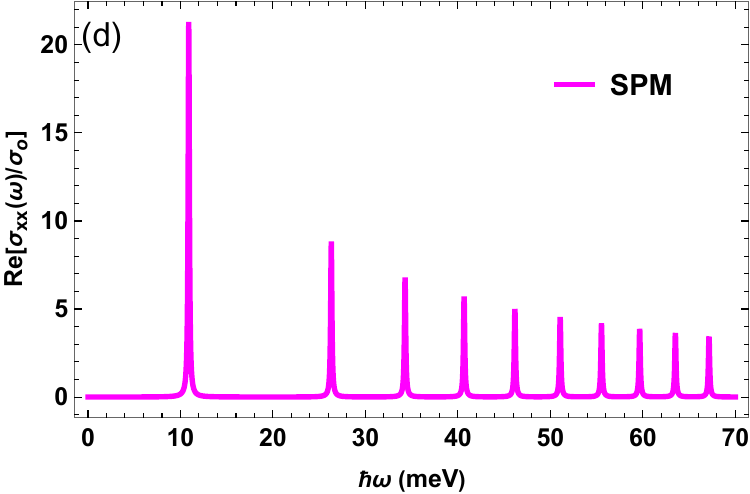}\\
	\includegraphics[width=0.450\linewidth]{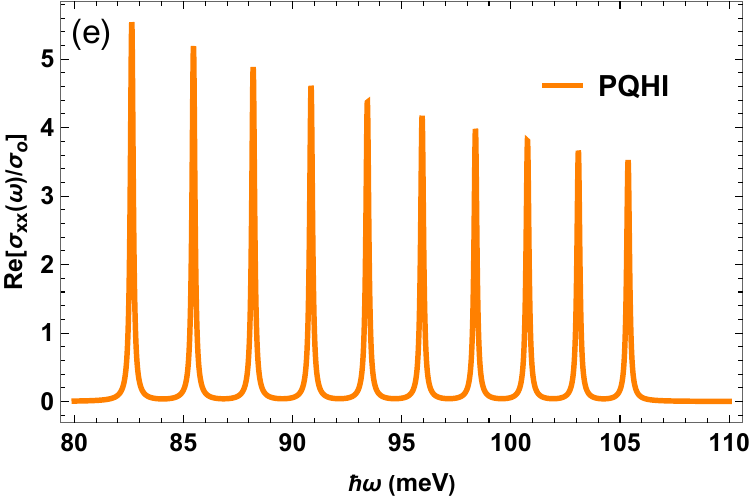}
	\includegraphics[width=0.450\linewidth]{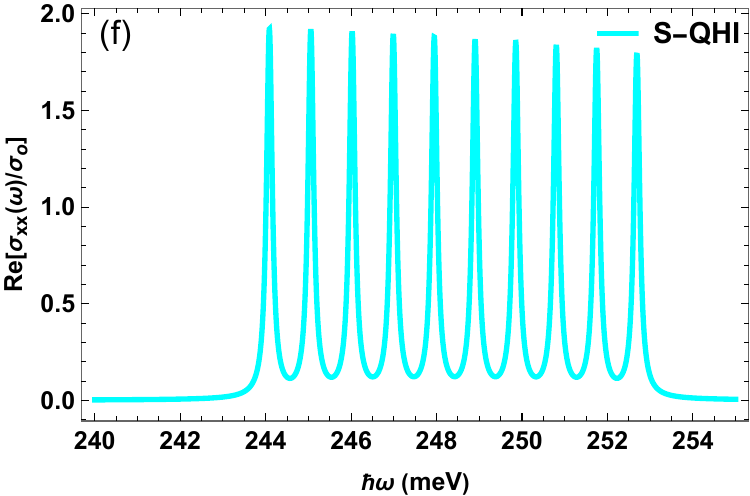}
	\caption{The real components of $\sigma_{xx}$ as a function of photon energies (in units of meV) in different topological phases for spin-down electrons in the $K$ valley. (a) QSHI ($V=0.5\lambda_{so}$ $\lambda_{\omega}=0$), (b) VSPM  ($V=\lambda_{so}$ $\lambda_{\omega}=0$), (c) BI ($V=1.5\lambda_{so}$,  $\lambda_{\omega}=0$), (d) SPM ($V=0$,  $\lambda_{\omega}=\lambda_{so}$), (e) PQHI ($V=0$, $\lambda_{\omega}=1.5\lambda_{so}$) and (f) S-QHI ($V=\lambda_{so}$,  $\lambda_{\omega}=1.5\lambda_{so}$) respectively. The corresponding LLs to these conductivities are shown in Figs.~\ref{LL1}(a)--(f), respectively.}
	\label{Sigma2}
\end{figure*}

\begin{figure*}[ht!]
	\centering		
	\includegraphics[width=0.450\linewidth]{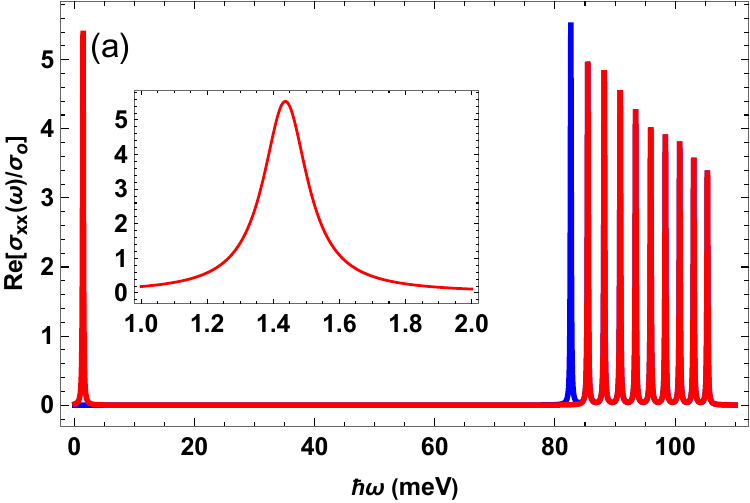}
	\includegraphics[width=0.450\linewidth]{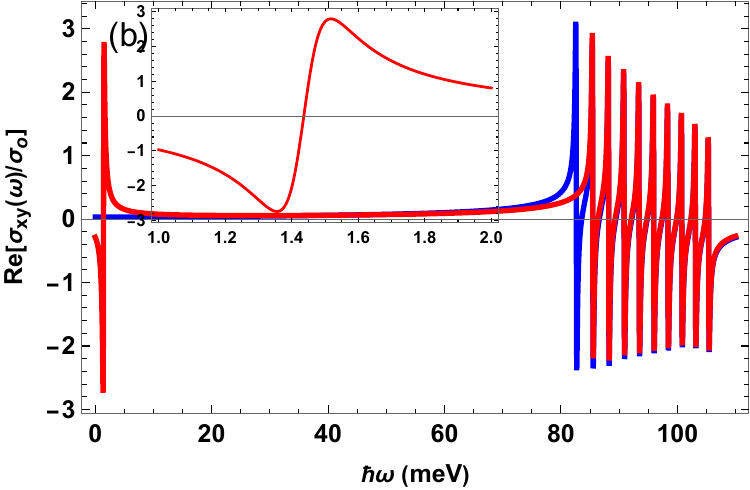}
	\caption{The real components of the (a) $\sigma_{xx}$ and (b) $\sigma_{xy}$ of ML-jacutingaite as a function of photon energies in the QSHI ($V=0.5\lambda_{so}$ $\lambda_{\omega}=0$) phase and fixed magnetic field $B=1$ for spin-down electrons in the $K$ valley. The chemical potential is laying in between $n=0$ and $n=1$ ($\mu_{F}=42$ meV). The Pauli-blocked transitions are shown by blue solid curves, while the curve in the inset represents the allowed intraband optical transition originating instead of the Pauli-blocked interband transition.}
	\label{Sigma33}
\end{figure*}
The Kronecker delta functions in these expressions guarantee that electric dipole selection rules for LL transitions are fulfilled. On the other hand, the Heaviside step functions \( \Theta(E_n - \mu_F) \) restrict the transitions to those crossing the Fermi level, thereby accounting for Pauli blocking effects \cite{grigorenko2012graphene}. Similarly, we can express the real and imaginary parts of the transverse Hall conductivity  \cite{shah2019magneto}
\begin{equation}\label{a18}
\begin{aligned}
\operatorname{Re} & \bigg(\frac{\sigma_{x y}(\omega)}{\sigma_0}\bigg)=\frac{2 v^2 \hbar e B}{\pi} \sum_{\eta, s= \pm} \sum_{n, m=0}^{\infty} \sum_{t, t^{\prime}= \pm} \eta \frac{\Theta\left(\mu-E_{m, t'}^{\eta, s}\right)-\Theta\left(\mu-E_{n, t}^{\eta, s}\right)}{E_{n, t}^{\eta, s}-E_{n, t'}^{\eta, s}} \\
& \times\left[\left(\mathcal{A}_{m, t'}^{\eta, s} \mathcal{B}_{n, t}^{\eta, s}\right)^2 \delta_{n, m-\eta}-\left(\mathcal{B}_{m, t'}^{\eta, s} \mathcal{A}_{n, t}^{\eta, s}\right)^2 \delta_{n, m+\eta}\right] \frac{\hbar \omega+E_{m, t'}^{\eta, s}-E_{n, t}^{\eta, s}}{\Gamma^2+\left(\hbar \omega+E_{m, t'}^{\eta, s}-E_{n, t}^{\eta, s}\right)^2},
\end{aligned}
\end{equation}
and
\begin{equation}\label{a19}
\begin{aligned}
\operatorname{Im} & \bigg(\frac{\sigma_{x y}(\omega)}{\sigma_0}\bigg)=-\frac{2 v^2 \hbar e B}{\pi} \sum_{\eta, s= \pm n, m=0} \sum_{t, t^{\prime}= \pm}^{\infty} \sum_n \eta \frac{\Theta\left(\mu_{F}-E_{m, t'}^{\eta, s}\right)-\Theta\left(\mu_{F}-E_{n, t}^{\eta, s}\right)}{E_{n, t}^{\eta, s}-E_{m, t'}^{\eta, s}} \\
& \times\left[\left(\mathcal{A}_{m, t'}^{\eta, s} \mathcal{B}_{n, t}^{\eta, s}\right)^2 \delta_{n, m-\eta}-\left(\mathcal{B}_{m, t'}^{\eta, s} \mathcal{A}_{n, t}^{\eta, s}\right)^2 \delta_{n, m+\eta}\right] \frac{\Gamma}{\Gamma^2+\left(\hbar \omega+E_{m, t'}^{\eta, s}-E_{n, t}^{\eta, s}\right)^2}.
\end{aligned}
\end{equation}
\end{widetext}
In the limit where $\Delta_{so} = 0$, $\lambda_{\omega}=0$ and $V = 0$, the system recovers the unconventional Hall conductivity behavior characteristic of graphene \cite{gusynin2005unconventional}. In these expressions, the real (imaginary) component of \( \sigma_{xx} \) (\( \sigma_{xy} \)) consists of a sum of absorptive Lorentzian functions, where each peak’s full width at half maximum (FWHM) is determined by the scattering rate \( \Gamma \); increasing \( \Gamma \) leads to broader and less intense peaks. Similarly, the real (imaginary) component of \( \sigma_{xy} \) (\( \sigma_{xx} \)) is composed of dispersive Lorentzian functions, with peaks located at the magneto-excitation energies \( \hbar\omega = E_n - E_m \). These optical transitions obey strict selection rules: \( |n| - |m| = \pm 1 \), along with spin conservation. As a result, transitions between states of opposite spin (\( \sigma = +1 \) and \( \sigma = -1 \)) are spin-forbidden.

To gain a comprehensive understanding of the magneto-optical conductivity of ML-jacutingaite, we present the LLs as a function of the magnetic field \( B \), along with the allowed LL transitions, in Fig.~\ref{LL1_graphene}. The corresponding magneto-optical excitation energies associated with these transitions, given by \( \mathcal{E}^{\eta,s}_{m} - \mathcal{E}^{\eta,s}_{n} \), are labeled as \( T^{\eta,s}_{m \rightarrow n} \).  Figures~\ref{Sigma1}(a) and (b) display the real part of $\sigma_{xx}$ as a function of $\hbar\omega$ for spin-up and spin-down states in the $K$ valley, respectively, within QSHI phase.

The longitudinal conductivity $\sigma_{xx}$ spectra exhibit multiple absorption peaks at different magneto-excitation energies, corresponding to electron transitions between different LLs for both spin-up and spin-down states. Resonant peaks in the magneto-optical conductivity spectra arise when the photon energy matches the energy gap between LLs for a given magnetic field. These peaks manifest as absorption features in optical spectroscopy and could potentially be used to extract the band parameters or identify the topological states. In Figs.~\ref{Sigma1}(a) and (b), we can observe multiple absorption peaks in the longitudinal conductivity spectra for both spin-up and spin-down electrons, respectively. In this analysis, we focus exclusively on transitions between the $n=0\rightarrow n=1$ and $n=1\rightarrow n=2$ LLs. For example, in Fig.~\ref{Sigma1}(a) the two main peaks of $\sigma_{xx}$  spectra correspond to the transitions $T^{K,\downarrow}_{-1\rightarrow 0}$ and $T^{K,\downarrow}_{-1\rightarrow 2}$ for spin-down in the $K$ valley for different magnetic fields. The other conductivity peaks located at higher
photonic energies correspond to electron transitions between higher LLs, as illustrated in Fig.~\ref{LL1_graphene}(a). It is evident that the transition \( T^{K,\downarrow}_{-1 \rightarrow 0} \) exhibits significantly stronger intensity compared to other LL transitions, dominating the conductivity spectrum. As the magnetic field \( B \) increases, the corresponding magneto-excitation energies shift to higher photon energies. This results in a noticeable shift in the resonant peak of the \( \sigma_{xx} \) spectrum, as illustrated in Fig.~\ref{Sigma1}(a). Additionally, Fig.~\ref{LL1_graphene}(b) depicts the LLs and allowed transitions for spin-up electrons in the \( K \) valley of ML-jacutingaite. In contrast to the spin-down case, spin-up electrons experience a much larger bandgap, leading to magneto-optical transitions occurring at higher photon energies. Next, we calculate the real part of \( \sigma_{xx} \) vs the photon energy, and the results are shown in Figs.~\ref{Sigma1}(c) and (d), corresponding to spin-down and spin-up states, respectively. The number of paired positive-negative peaks in the real part of $\sigma_{xy}$ directly corresponds to the absorption features in the real part of $\sigma_{xx}$.

Figures~\ref{Sigma2}(a)--(f) shows the magneto-optical longitudinal conductivity as a function of incident photon energy across various topological phases. We plotted $\sigma_{xx}$ for the spin-down electrons in the $K$ valley for a magnetic field of 1 T. The presence of a finite staggered potential induces spin-dependent splitting of the LLs, leading to distinct energy separations for the spin-down states. This splitting causes an uneven distribution of spectral weight between the spin-resolved transitions, thereby diminishing the intensity of the inter-band absorption peaks as shown in Fig.~\ref{Sigma2}(a). In the case corresponding to the VSPM phase—spin-down absorption peaks shift to lower photon energies (redshift), indicating the closure of the smallest band gap for Dirac fermions (see Fig.~\ref{Sigma2}(b)). Conversely, in the BI phase, the absorption peaks shift to higher energies due to the widening of the band gap as illustrated in Fig.~\ref{Sigma2}(c). when $V=0$ and $\lambda_\omega$ increases, the system transitions from the QSHI to the SPM phase. In this regime, the closing of the band gap results in all inter-band absorption features shifting toward lower photon energies, as shown in Fig.~\ref{Sigma2}(d). For finite values of $V$ and $\lambda_\omega$, the system enters the PQHI regime. In this phase, the absorption peaks in the real part of the longitudinal conductivity spectra exhibit behavior similar to that observed in the QSHI phase as shown in Fig.~\ref{Sigma2}(e). In the S-QHI phase, the spacing between spin-down electron Ls becomes more pronounced as depicted in Fig.~\ref{LL1}(f). As a result, the corresponding absorption peaks appear at higher photon energies in the conductivity spectra as showcased in Fig.~\ref{Sigma2}(f).

The magneto-optical conductivities of ML-jacutingaite are strongly dependent on the chemical potential $\mu_{F}$. This potential can be tuned via external bias or optical pumping, altering electron densities and thereby influencing ML-jacutingaite's properties. Crucially, interband transitions occur between LLs (LLs) from different bands, while intraband transitions occur between LLs within the same band; these distinct processes affect the material's optical properties differently. Figs. ~\ref{Sigma33}(a) and (b) represent the real components of the longitudinal and transverse conductivities of ML-jacutingaite as a func-
tion of photon energies with $\mu_{F}$ = 42 meV in the QSHI phase and fixed magnetic field $B$ = 1 for spin-down electrons in the $K$ valley.  The chemical potential lies between $n =0$  and $n =1$ LL.  With $\mu_{F}$ = 42 meV, certain LLs transitions are Pauli-blocked due to occupied states. For instance,  the transition $T^{K,\downarrow}_{-1\rightarrow 0}$=82.6 meV becomes Pauli-blocked, and the intraband transition $T^{K,\downarrow}_{0\rightarrow 1}=1.4$ meV emerges in its place. For $\mu_{F}$ = 42 meV, we can see absorptive intraband and interband features for the lowest energy transitions for the spin-down electrons, respectively. The Pauli-blocked magneto-optical transitions are shown by blue solid curves, while the curve in the inset represents the allowed intraband optical transition originating instead of the Pauli-blocked interband transition.

\section{conclusion}
 We have theoretically explored quantum magnetotransport in a monolayer jacutingaite 2D system in the presence of electric fields, magnetic fields, and off-resonant circularly polarized light. Our study demonstrates a diverse range of topological phases, each defined by variations in the spin- and valley-resolved Chern numbers, which are influenced by external factors such as optical field intensities and staggered sublattice potentials. 

 The zeroth LL in jacutingaite has exhibited distinct peaks in the DOSs. We have calculated both longitudinal and Hall magneto-optical conductivities using the Kubo formalism and showed how their spectral features are highly tunable by external fields and doping. The optical transitions between LLs exhibit resonance peaks that serve as signatures of underlying topological phases.

 Our findings demonstrate that monolayer jacutingaite is an exceptionally manageable two-dimensional quantum material with valuable potential uses in future advancements of topological electronics, valleytronics, and optospintronics. The ability to adjust its band structure and optical behavior positions it as an excellent option for practical investigation and use in quantum technology systems.

\renewcommand{\bibname}{References}

\section*{References}

\bibliography{refr}
\end{document}